\documentclass[10pt,twocolumn,letterpaper]{article}

\usepackage[pagenumbers]{cvpr}

\newcommand*{\ShowNotes}{} 

\definecolor{noteone}{HTML}{DBB053}
\definecolor{notetwo}{HTML}{60DBA6}
\definecolor{notethree}{HTML}{BA60DB}

\ifdefined\ShowNotes
  \newcommand{\colornote}[3]{{\textbf{\color{#1}[#2 #3]}}}
\else
  \newcommand{\colornote}[3]{}
\fi

\newcommand{\method}{\textsc{Copy-Transform-Paste}}

\usepackage[export]{adjustbox}
\usepackage{xcolor,colortbl}
\usepackage{multirow}
\usepackage{pifont}
\usepackage{float}
\usepackage{caption}
\usepackage{tikz}
\usepackage{relsize}
\usepackage[outline]{contour}
\contourlength{1.2pt}
\usepackage{amssymb}    
\usepackage{graphicx}
\usepackage{amsmath}
\usepackage{booktabs}
\usepackage{siunitx}

\definecolor{cvprblue}{rgb}{0.21,0.49,0.74}
\usepackage[pagebackref,breaklinks,colorlinks,allcolors=cvprblue]{hyperref}

\title{\method: Zero-Shot Object-Object Alignment Guided by Vision-Language and Geometric Constraints}

\def\authorBlock{
    Rotem Gatenyo \qquad
    Ohad Fried \qquad
    \\[0.4em]
    Reichman University\\
    {\tt\small \href{https://rotemgat.github.io/CopyTransformPaste/}{\color{magenta}{GitHub Page}}}
}
\author{\authorBlock}

\begin{document}

\twocolumn[{%
\renewcommand\twocolumn[1][]{#1}%
\maketitle
\begin{center}
    \centering
    \captionsetup{type=figure}
    \includegraphics[width=\textwidth]{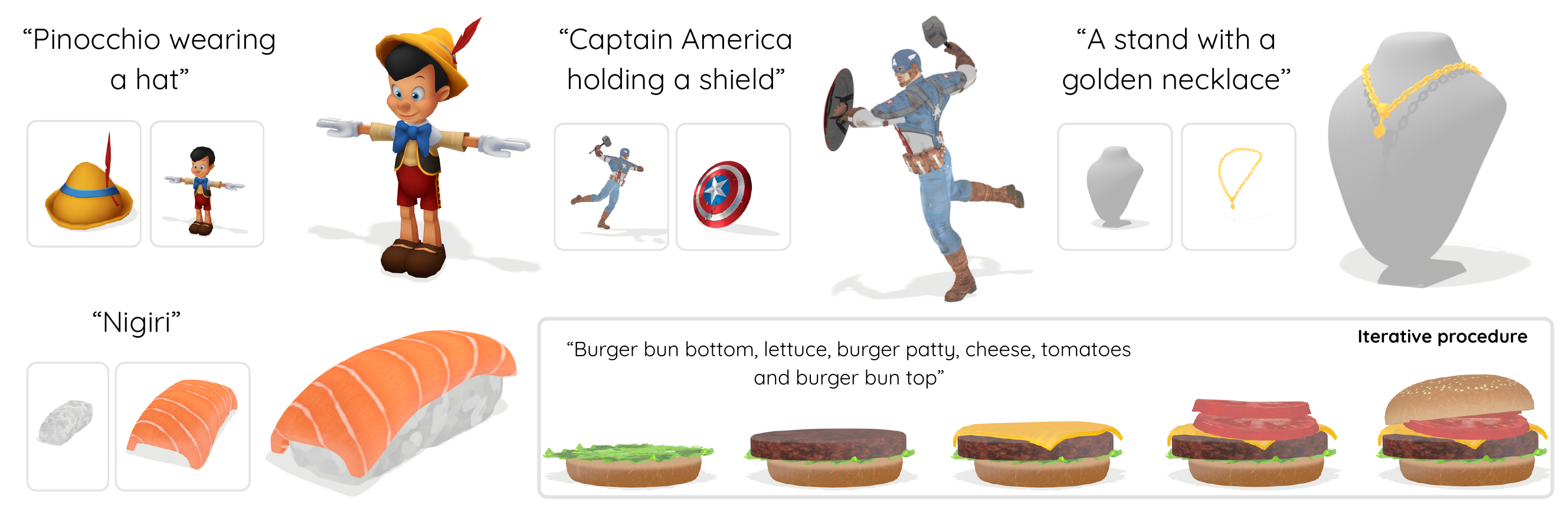}
    \captionof{figure}{
    \textbf{Text-guided object-object alignment and iterative composition.}
    The figure shows four independent examples, each presenting the input meshes and text prompt alongside our alignment result.
    In addition, an iterative example demonstrates progressive assembly of a burger: the output of stage \(k\) is incorporated into the input of stage \(k{+}1\), gradually forming the final arrangement.}
    \label{fig:teaser}
\end{center}%
}]

\begin{abstract}
We study zero-shot 3D alignment of two given meshes, using a text prompt describing their spatial relation---an essential capability for content creation and scene assembly. Earlier approaches primarily rely on geometric alignment procedures, while recent work leverages pretrained 2D diffusion models to model language-conditioned object-object spatial relationships. In contrast, we directly optimize the relative pose at test time, updating translation, rotation, and isotropic scale with CLIP-driven gradients via a differentiable renderer, without training a new model.
Our framework augments language supervision with geometry-aware objectives: a variant of soft-Iterative Closest Point (ICP) term to encourage surface attachment and a penetration loss to discourage interpenetration. A phased schedule strengthens contact constraints over time, and camera control concentrates the optimization on the interaction region.
To enable evaluation, we curate a benchmark containing diverse categories and relations, and compare against baselines. 
Our method outperforms all alternatives, yielding semantically faithful and physically plausible alignments.
\end{abstract}
\vspace{-1.2em}
\section{Introduction}
\label{sec:intro}

Objects rarely exist in isolation: everyday 3D tasks such as placing a cup on a saucer, fitting a lid on a pot or topping a sundae with a cherry, require arranging one object relative to another in a way that is both semantically intended and physically plausible. We focus on the fundamental problem of aligning two given meshes according to a short text prompt that guides the intended interaction.

Data scarcity makes this problem challenging. Unlike human-object interaction (HOI), which benefits from contact-rich datasets and evaluation protocols~\cite{taheri2020grab,bhatnagar2022behave}, object-object interaction lacks comparable resources. Large-scale, standardized benchmarks remain limited; to our knowledge, 2BY2~\cite{qi2025two} is the most extensive to date, yet it covers only 18 pairwise alignment tasks. This motivates a zero-shot approach that relies on pretrained models at test time rather than supervision on 3D alignment data.

To leverage vision-language signals in 3D, we expose pose parameters to image-space objectives via differentiable rendering~\cite{kato2018neural,liu2019soft,laine2020modular}. 
Differentiable rendering provides gradients through rendering, enabling direct optimization of explicit mesh parameters from rendered supervision~\cite{gao2023textdeformer,chen2024text}.

We use CLIP~\cite{radford2021learning} as the language-vision objective: CLIP embeds text and images in a joint space, allowing us to measure cosine similarity between rendered views and the prompt. Prior 3D works have shown that CLIP supervision can drive mesh editing and stylization through rendered views~\cite{michel2022text2mesh,gao2023textdeformer, mohammad2022clip}; we adopt the same supervision to inform 3D placement.

Language supervision alone does not impose contact or prevent interpenetration. Classical alignment offers complementary structure: Iterative Closest Point (ICP) aligns two shapes by alternating nearest-neighbor correspondence with a rigid pose update that minimizes point-to-point distances~\cite{besl1992method}, with robust/probabilistic variants improving tolerance to noise and partial overlap~\cite{chen1992object,myronenko2010point}. Building on these ideas, we introduce a \emph{fractional soft-ICP} attachment term, our variant that applies soft correspondences only to a prescribed fraction of the closest vertices, to encourage controlled surface contact, and we add a penetration penalty to discourage interpenetration.

At test time, we perform a phased optimization that progressively increases the weights of the fractional soft-ICP and penetration terms while zooming cameras toward the interaction region; multiple random restarts with best-of-\(N\) selection improve robustness to initialization. We validate on a curated benchmark of 50 mesh-prompt pairs against geometric and LLM-based baselines, observing higher semantic agreement and lower intersection volume.

\noindent Our main \textbf{contributions} are:
\begin{enumerate}[leftmargin=*,nosep]
\item A text-guided, test-time optimization framework that estimates the relative pose and isotropic scale between two meshes via differentiable rendering and vision-language supervision, augmented with fractional soft-ICP and penetration objectives for physical plausibility. 
\item A benchmark of 50 language-conditioned mesh-pair cases for standardized evaluation of object-object alignment (OOA), against which we demonstrate consistent improvements over multiple baselines. 
\end{enumerate}

\section{Related Work}
\label{sec:related_work}

\noindent\textbf{Classical alignment and mesh composition.}
Rigid alignment of two shapes is commonly solved with ICP: starting from an initial pose, the algorithm alternates (i) establishing correspondences between samples on the source and target and (ii) estimating the rigid transform that best fits those correspondences, iterating to convergence~\cite{besl1992method,chen1992object}. Other ICP formulations cast correspondence as a probabilistic assignment, such that each source sample matches multiple targets with weights; this increases robustness to noise, partial overlap, and outliers~\cite{chen1992object,myronenko2010point}. Building on these primitives, interactive composition systems apply ICP within user-guided pipelines to assemble parts and enforce contact; a representative example is SnapPaste~\cite{sharf2006snappaste}, which performs cut-and-paste composition with contact-aware snapping. Beyond these representatives, there is a large body of ICP variants and comparisons; see, e.g., ~\citet{Rusinkiewicz2001EfficientVO} and \citet{tam2012registration} for analyses of efficient ICP variants and 
3D registration. These methods focus on geometric consistency and interaction cues, but do not inject semantic guidance from language or images.

\vspace{0.55em}
\noindent\textbf{Text-guided optimization with vision-language models.}
A large body of text-guided content synthesis is driven by CLIP~\cite{radford2021learning}, a foundational model that learns a joint embedding space for text and images. Optimization under CLIP guidance has proven to be effective for semantic control in images~\cite{patashnik2021styleclip,kim2022diffusionclip,gal2022stylegan} and vector sketches~\cite{vinker2022clipasso,frans2022clipdraw,song2023clipvg}.
In 3D, text-driven optimization has been used to steer shape generation directly from language~\cite{poole2022dreamfusion, jain2022zero}. These methods demonstrate that language priors can guide optimization toward semantically meaningful configurations.

\vspace{0.55em}
\noindent\textbf{Optimization on 3D meshes with differentiable rendering.}
Other works adopt surface-based differentiable rendering to render explicit meshes and pass those views to vision-language models, enabling gradients from language-image objectives to propagate back to 3D parameters~\cite{kato2018neural,liu2019soft,laine2020modular}. Within this setup, Text2Mesh~\cite{michel2022text2mesh} performs text-driven stylization of mesh appearance and local geometry, TextDeformer~\cite{gao2023textdeformer} manipulates mesh geometry under text guidance, and recent editing methods such as~\mbox{\citet{chen2024text}} and PrEditor3D~\cite{erkocc2025preditor3d} support controlled, localized edits. These advances can be leveraged for object-object interaction, where the goal is not to alter geometry but to recover a plausible pose between two meshes under a text prompt.

\vspace{0.55em}
\noindent In contrast to pipelines that rely purely on geometry and to language-driven editing methods that modify shape or appearance, we address zero-shot rigid pose and scale alignment between a given pair of meshes from a text prompt, coupling differentiable rendering-based language supervision with explicit contact and penetration objectives.

\section{Method}
\label{sec:method}

\begin{figure*}[t]
  \centering
  \includegraphics[width=\linewidth]{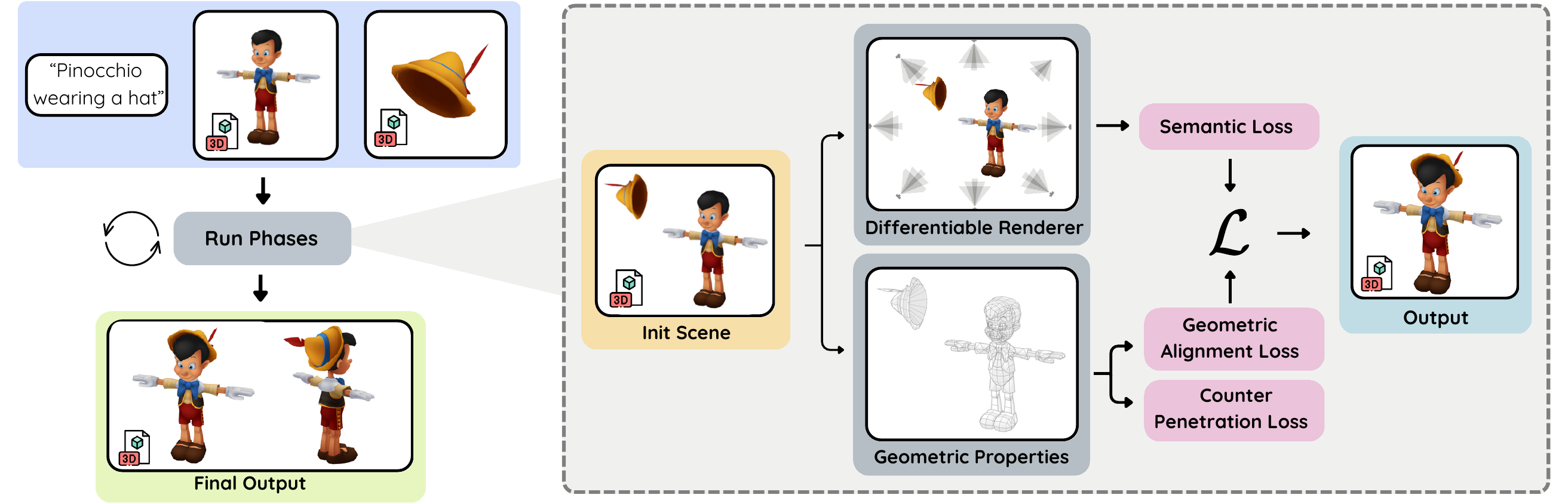}
  \caption{\textbf{Overview of the proposed pipeline.} Given two meshes and a text prompt, we optimize the relative pose and scale to produce a text-consistent alignment over $P$ phases. In each phase, we compose the scene, render with a differentiable renderer to obtain a semantic loss, and compute geometric losses.
  The best result of phase \(i\) initializes phase \(i{+}1\); across phases we increase the fractional soft-ICP and penetration weights and progressively zoom the cameras in. The final output is an aligned 3D placement of the two meshes.
  }
  \label{fig:method}
\end{figure*}

Our objective is to align two given meshes, guided by a textual description $t$ that provides semantic context for the interaction. We refer to the two meshes as \emph{source} and \emph{target}; this designation is arbitrary, either mesh can play either role. The method optimizes the pose parameters $\theta=(\tau,q,s)$; translation $\tau$, rotation $q$ (unit quaternion), and isotropic scale $s$ of the source mesh relative to the target until the rendered configuration agrees with $t$ while respecting geometric and physical constraints. An overview of the pipeline is illustrated in~\cref{fig:method}.

\subsection{Initialization}
We canonicalize the target mesh with Auto-Align~\cite{autoalign}, which estimates bilateral symmetry and reorients the mesh to a common upright frame~\cite{mitra2013symmetry}. This makes the rendered views consistent with the object’s typical orientation (\eg., a cup becomes upright rather than tilted), reducing view ambiguity and stabilizing semantic supervision. 

\subsection{Text Guidance}
To guide the pose toward alignment with a textual prompt, we define a semantic objective that encourages agreement between the prompt and the rendered views in a shared image-text embedding space. We use CLIP~\cite{radford2021learning} to provide this space by mapping images and text into joint embeddings. We render views with a differentiable renderer~\cite{laine2020modular} $\mathcal{R}$, which links geometry and images and enables gradients from the similarity score to backpropagate to the 3D pose parameters.

Let \(M_S\) and \(M_T\) be the source and target meshes, respectively. At each step, we update the pose parameters \(\theta\) of \(M_S\) and denote the posed mesh by \(\tilde{M}_S\). The scene is $S = M_T \cup \tilde{M}_S$.
We render \(N\) views \(\{I_i\}_{i=1}^{N}\) with the differentiable renderer \(\mathcal{R}\) from cameras \(\{c_i\}\) and encode each view and the prompt with CLIP, \(e_i=\texttt{CLIP}_{\text{img}}(I_i)\) and \(e_t=\texttt{CLIP}_{\text{text}}(t)\), and define the text-guidance loss
\begin{equation}
\label{eq:clip}
\mathcal{L}_{\text{clip}}
= -\frac{1}{N}\sum_{i=1}^{N}\mathrm{sim}\!\big(e_i,\,e_t\big),
\quad
\mathrm{sim}(e_i, e_t)
= \frac{e_i \cdot e_t}{\|e_i\|\,\|e_t\|}.
\end{equation}
See \cref{fig:init_prompt}b for a demonstration of prompt-driven controllability, 
where changing the text alone steers the final pose.

\subsection{Geometric Objectives}
Text guidance alone can yield semantically plausible but physically invalid placements. Two geometry-aware terms promote physical plausibility: (i) a fractional soft-ICP attachment term that pulls a controlled subset of source vertices toward the target vertices, and (ii) an interpenetration penalty that discourages collisions.

\paragraph{Fractional Soft ICP.}
Soft ICP replaces hard nearest-neighbor matches with probabilistic correspondences: each source vertex assigns a distribution over target vertices (\eg., Gaussian weights), and the pose update minimizes the expected squared distance under these weights, improving stability under noise and partial overlap~\cite{myronenko2010point}. 

\begin{figure}[t]
  \centering
  \includegraphics[width=\linewidth]{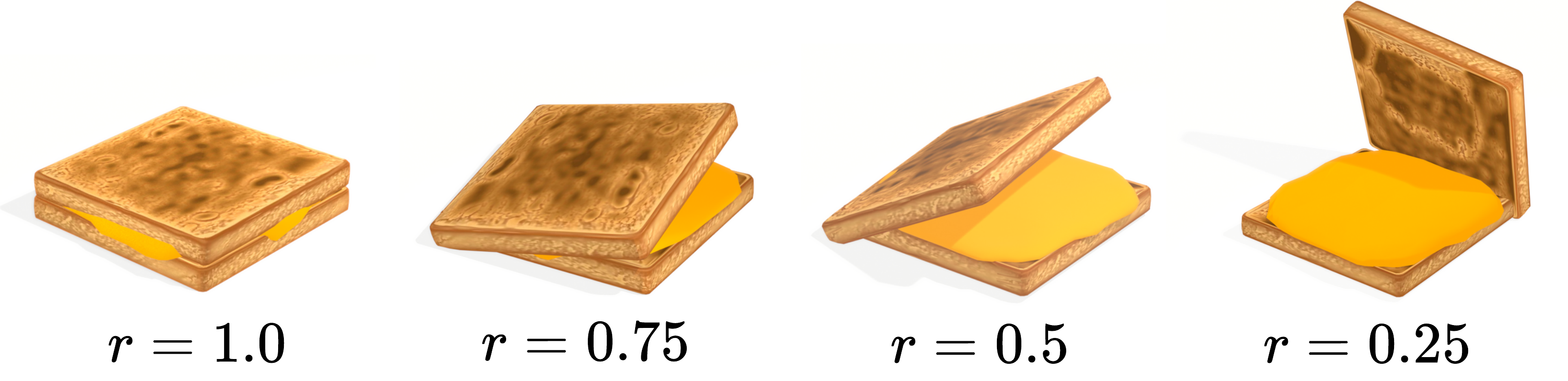}
  \caption{Effect of the alignment ratio $r$ on a grilled-toast pair. 
  The two objects are optimized with the same prompt, ``grilled cheese toasts'', while varying $r$. 
  With $r{=}1.0$, the top toast aligns directly above the bottom toast, producing broad surface contact; as $r$ decreases, attachment is encouraged over a smaller subset of vertices and the contact region correspondingly shrinks. 
  }
  \label{fig:icp_fraction}
\end{figure}

We introduce a \emph{fractional} variant that limits attachment to only the closest subset of source vertices, controlled by a ratio \(r\in(0,1]\). Let \(V_S=\{v_i^S\}_{i=1}^{N_S}\) and \(V_T=\{v_j^T\}_{j=1}^{N_T}\) be the vertex sets of the source and target meshes. For each \(v_i^S\), compute its nearest-target distance \(d_i^{\min}=\min_j\|v_i^S-v_j^T\|_2\). Select the \(K=\lfloor rN_S\rfloor\) indices with the smallest \(d_i^{\min}\) to form \(W\). Define squared distances \(E_{ij}=\|v_i^S-v_j^T\|_2^2\) and soft correspondences by a softmax \emph{over targets \(j\)}:
\[
\alpha_{ij}=\frac{\exp\!\big(-E_{ij}/(2\sigma^2)\big)}{\sum_{j'=1}^{N_T}\exp\!\big(-E_{ij'}/(2\sigma^2)\big)},\qquad \sum_{j}\alpha_{ij}=1.
\]
The attachment loss averages the expected squared distance over the selected subset:
\begin{equation}
\label{eq:icp}
\mathcal{L}_{\text{icp}}(r)=\frac{1}{K}\sum_{i\in W}\sum_{j=1}^{N_T}\alpha_{ij}\,E_{ij}.
\end{equation}
Unlike standard soft-ICP, which uses all source vertices, \emph{fractional} soft-ICP attaches only the closest \(r\)-fraction (recovering soft-ICP when \(r{=}1\)). \cref{fig:icp_fraction} visualizes the effect of the alignment ratio \(r\): larger \(r\) enforces broader attachment across contacting surfaces, while smaller \(r\) limits attachment to a narrower region.

\begin{figure}[t]
  \centering
  \includegraphics[width=\columnwidth]{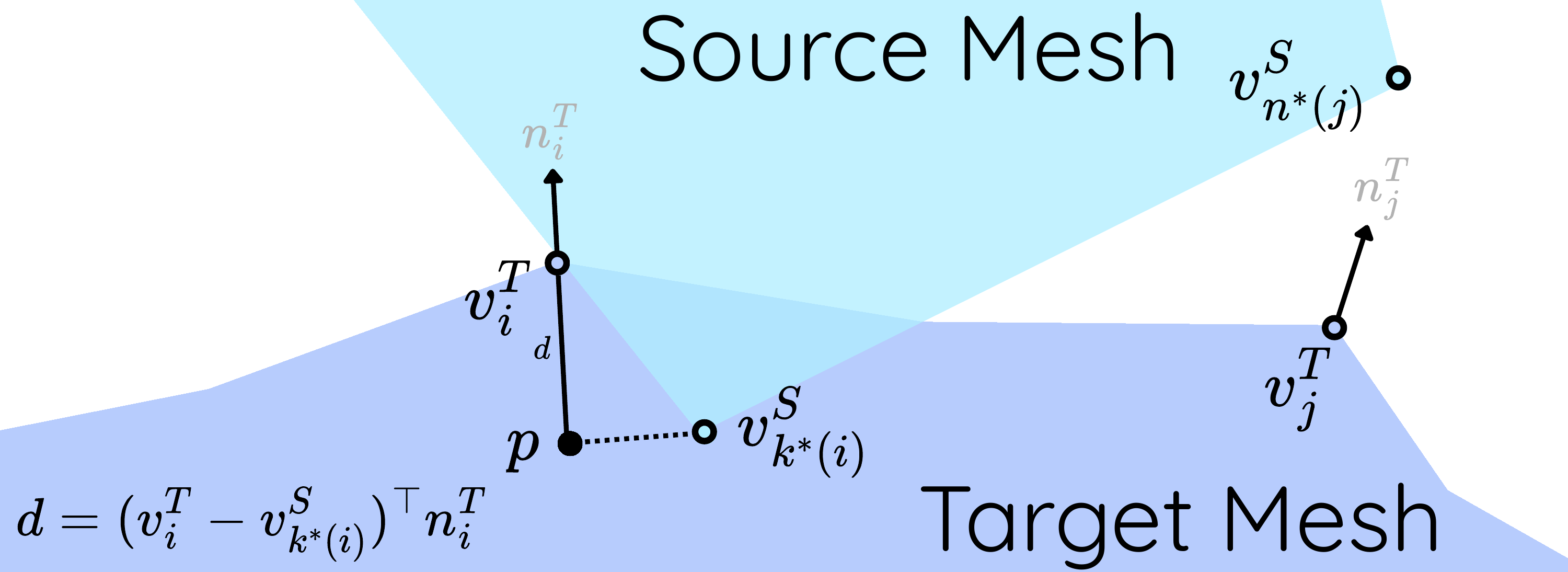}
  \caption{\textbf{Penetration loss geometry.}
  For target vertex $v_j^T$, the nearest source vertex lies outside the surface, so the signed depth and the loss term is zero.
  For target vertex $v_i^T$, the nearest source vertex is inside; the orthogonal projection onto the normal yields point $p$, and the signed depth $d_i>0$ produces a positive penalty.}
  \label{fig:penetration_loss}
\end{figure}

\paragraph{Penetration Loss.}
The penetration loss prevents invalid overlap by penalizing intrusion of the source mesh into the target mesh; a small positive margin permits limited indentation for soft materials.  
As in ContactOpt~\cite{grady2021contactopt}, penetration beyond a margin $c_{\text{pen}}$ is penalized via signed depth along outward normals of the target surface. Let $\{(v_j^T, n_j^T)\}_{j=1}^{N_T}$ denote pairs of target-surface vertices and their outward normals, and let $v_{i^*(j)}^S$ be the Euclidean nearest source vertex to $v_j^T$:
\begin{equation}
\label{eq:intersect}
\mathcal{L}_{\text{pen}}
\;=\;
\sum_{j=1}^{N_T}
\max\!\left(0,\; (v_j^T - v_{i^*(j)}^S)^\top n_j^T - c_{\text{pen}}\right)
\end{equation}
The margin $c_{\text{pen}}$ governs admissible indentation: $c_{\text{pen}}{=}0$ for rigid contact; $c_{\text{pen}}{>}0$ (e.g., 2\,mm) for soft materials. See \cref{fig:penetration_loss} for an illustration.

\subsection{Optimization and Scheduling}
The full objective combines text guidance and geometric constraints:
\begin{equation}
\label{eq:total}
\mathcal{L}=
\lambda_{\text{CLIP}}\mathcal{L}_{\text{clip}}
+\lambda_{\text{ICP}}\mathcal{L}_{\text{icp}}
+\lambda_{\text{pen}}\mathcal{L}_{\text{pen}}
\end{equation}
The Adam optimizer updates $\tau$, $q$, and $s$ of ${M}_T$. Gradients for $\mathcal{L}_{\text{clip}}$ are obtained via the differentiable renderer, while $\mathcal{L}_{\text{icp}}$, and $\mathcal{L}_{\text{pen}}$ are computed directly from mesh geometry.

\begin{figure}[t]
  \centering
  \includegraphics[width=\linewidth]{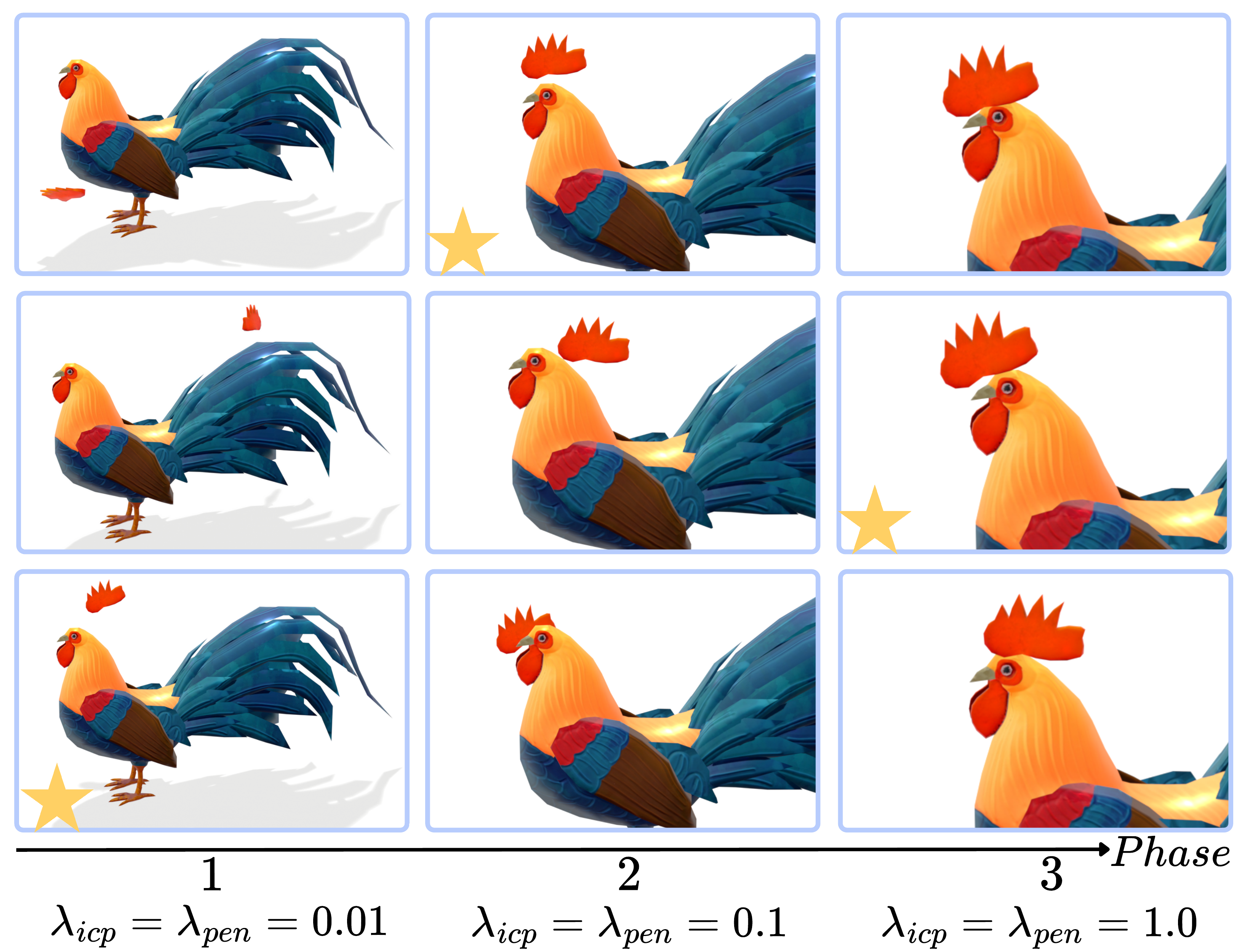}
  \caption{\textbf{Visualization of phased optimization with scheduled weights.}
  Rooster and comb across three phases. 
  As the weights increase across phases, the search transitions from broad exploration to a focused zoom-in and local refinement. 
  Phase-best results are marked with \(\star\) and initialize the next phase.
  }
  \label{fig:phase_schedule}
\end{figure}

\paragraph{Phased optimization}
To balance early exploration with later contact enforcement, optimization proceeds in $P$ successive phases. Each phase runs for a fixed number of steps; the best-scoring pose at the end of phase $p$ initializes phase $p{+}1$. Two schedules are applied: (i) the fractional soft-ICP weight is increased across phases, preventing premature sticking early on and consolidating attachment later; (ii) the penetration-loss weight is similarly increased, permitting limited passage through the target mesh in early phases (e.g., inserting a flower into a vase) and progressively suppressing interpenetration afterward. Early phases emphasize exploration of candidate contact regions under language guidance with relaxed attachment and penetration; later phases focus on refinement---settling on a region, reducing interpenetration, and fine-tuning the pose. ~\cref{fig:phase_schedule} visualizes this progression on a rooster-comb pair with
\(\lambda_{\text{ICP}}\) and \(\lambda_{\text{pen}}\) increased by \(\times 10\) between the three phases.

\paragraph{Camera scheduling}\label{subsec:camera}
Large scale disparities (e.g., a very small source mesh within a large target scene) can dilute gradients from language-vision scoring. Cameras are therefore scheduled across phases to progressively focus on the source mesh. Let $\mathbf{c}_t$ be the target-mesh centroid and $\mathbf{c}_s^{(p)}$ the source-mesh centroid at phase $p$. The per-phase look-at target interpolates from the target to the source center,
\[
\mathbf{c}^{(p)} \;=\; (1-\beta_p)\,\mathbf{c}_t \;+\; \beta_p\,\mathbf{c}_s^{(p)}, \quad 0=\beta_1<\cdots<\beta_P\le 1.
\]
Simultaneously, the camera distance is reduced to zoom in while keeping coverage. Early phases thus provide global context; later phases concentrate on the source mesh to expose finer details to the vision-language guidance.
See \cref{fig:phase_schedule} for an example of this zoom-in and look-at progression across phases.

\subsection{Random restarts and per-step noise}\label{subsec:random_restarts}
\begin{figure}[t]
  \centering
  \includegraphics[width=\columnwidth]{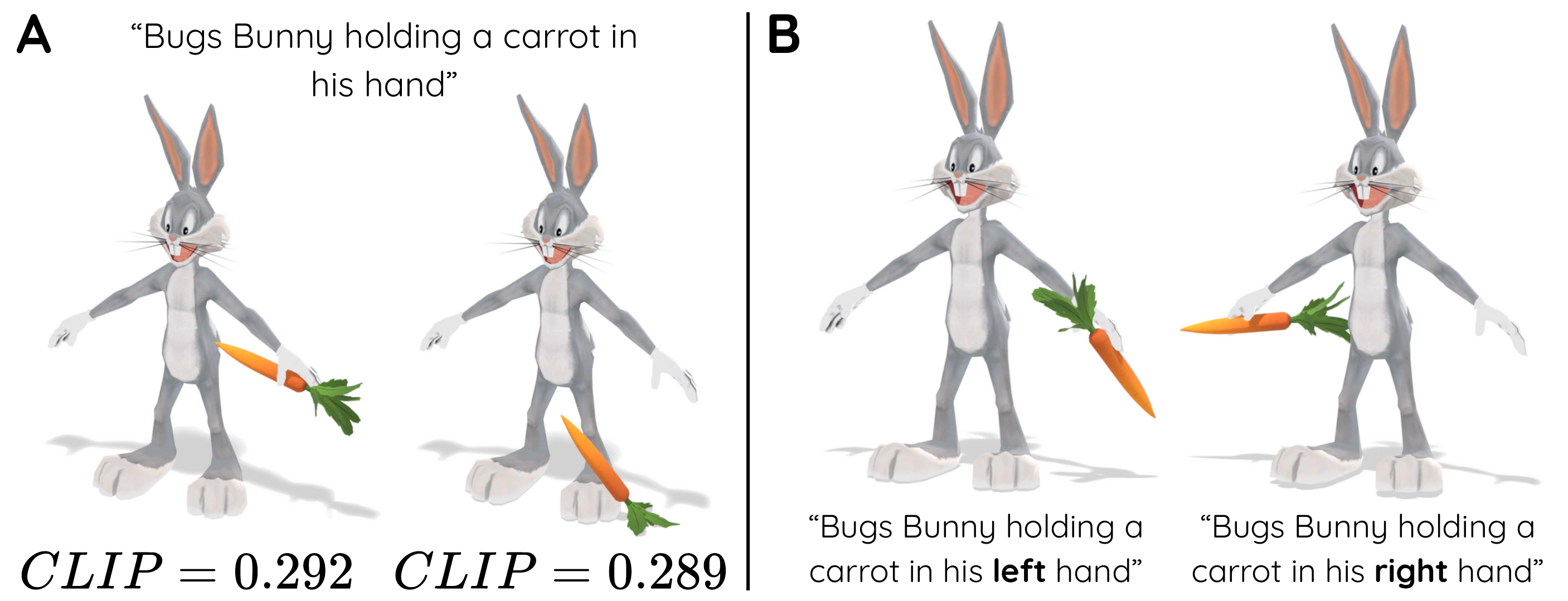}
  \caption{\textbf{Initialization variability and prompt controllability.}
  \textbf{(a)} Two random initializations of the carrot w.r.t.\ Bugs Bunny converge to distinct yet plausible attachments; we run several restarts and pick the best by a CLIP text–image score (higher is better). \textbf{(b)} With the same meshes, two prompts steer optimization to different, prompt-consistent placements, demonstrating language controllability.}
  \label{fig:init_prompt}
\end{figure}

Pose optimization is local and sensitive to initialization: if the source mesh starts near an unintended region of the target mesh, the optimizer can converge to a spurious contact. To mitigate this, the procedure launch $N$ independent initializations and runs the same objective for a fixed number of optimization steps. The final estimate is chosen by the highest evaluation score.  
\cref{fig:init_prompt}a illustrates the impact of initialization. 
Starting from two different random placements of the carrot relative to Bugs Bunny, 
the optimization converges to distinct yet plausible attachment points. 
This sensitivity motivates running multiple restarts and selecting the best result under the objective.

In addition, a small zero-mean perturbation is added to the pose parameters at each iteration. This stochastic jitter encourages exploration and helps escape local minima.

\subsection{LLM-guided hyperparameter selection}\label{subsec:hparam_llm}
At test time, we query a large language model (LLM) with the prompt $t$ and the two object names to obtain scene priors that set a few method hyperparameters. 

\textbf{Penetration policy.}
The LLM returns a boolean indicating whether the final arrangement should involve penetration (e.g., \emph{“knife cuts an apple”} $\Rightarrow$ allow penetration). If penetration is allowed, we set the penetration term weight to zero during optimization; otherwise we use the default penetration schedule.

\textbf{Initial scale.}
The LLM estimates the real-world size ratio between the source and target objects. We clamp the returned value to $[0.1,\,10]$ and use it to initialize the 
scale.

\textbf{Attachment ratio.}
For fractional soft-ICP, the LLM provides a coarse assessment of contact extent between the objects, which we map to the attachment ratio \(r \in [0,1]\) controlling the fraction of vertices selected for attachment.

\section{Experiments}
\label{sec:experiments}

\subsection{Experimental Setup}
\paragraph{Benchmark.}
Evaluation is conducted on a diverse set of 50 mesh pairs and text prompts covering a range of object-object relations, such as \textit{``A sundae with a cherry on top''}, \textit{``A candle sits inside a candle holder''} and \textit{``Pinocchio wearing a hat''}.
The benchmark, including meshes and prompts, will be released upon publication.

\paragraph{Computation and schedule.}
We run $2{,}000$ optimization steps per pair with a batch of $8$ camera views per step. 
Optimization proceeds in $P{=}3$ phases; across phases, the fractional soft-ICP and penetration weights follow a logarithmic increase, encouraging broad exploration early and stronger contact enforcement later. 
Camera scheduling (progressive zoom-in and look-at interpolation) follows \cref{subsec:camera}.

\paragraph{Initialization and Preprocessing.}
For meshes with low or uneven vertex density, we apply remeshing to ensure sufficient sampling for vertex-based geometric losses and to avoid degenerate boundary-only surfaces. To mitigate local minima, each experiment uses multiple randomized initial poses of the source mesh (see \cref{subsec:random_restarts}). 
We run \(N=5\) restarts under the same settings and automatically select the best-scoring pose by the total objective value.

\subsection{Metrics}
\label{sec:metrics}
OOA must satisfy both semantic intent and physical plausibility. We therefore report complementary semantic and geometric metrics.

\textbf{Semantic metrics.}
We render the final alignment from \(N{=}8\) fixed evaluation cameras and average image-text agreement across views using three vision-language encoders: CLIP~\cite{radford2021learning}, ALIGN~\cite{jia2021scaling}, and SigLIP~\cite{tschannen2025siglip}. Higher values indicate stronger text alignment.

\textbf{Intersection.}
Physical plausibility is assessed by volumetric overlap between the meshes. We report the \emph{intersection ratio}
\[
\mathrm{Inter.} \;=\; \frac{\mathrm{Vol}(\text{mesh}_1 \cap \text{mesh}_2)}{\mathrm{Vol}(\text{mesh}_1 \cup \text{mesh}_2)} \in [0,1].
\]
Lower values correspond to fewer inter-penetrations.

\textbf{VLM evaluator.}
We additionally report GPT-4V-based automatic scores following GPTEval3D~\cite{wu2024gpt},
covering four criteria: Text-Asset Alignment, 3D Plausibility,
Text-Geometry Alignment, and Overall. We omit the GPTEval3D Texture Details and Geometry Details criteria because all methods operate on the same input meshes; thus, texture and base geometry are shared (up to minor differences induced by the optimization), making these scores uninformative for method comparison.

\subsection{Baselines}
Language-guided OOA is a relatively new problem, thus methods with exactly the same objective are scarce. Accordingly, we evaluate against one method that pursues the same goal as ours, two closely related LLM-based methods, and two geometry-based methods.

\textbf{Object-object spatial relationships (OOR) diffusion}~\cite{baik2025learning}.
This method proceeds in two stages: (i) it creates training data by sampling a pretrained 2D diffusion model to synthesize diverse object-pair images and uplifting them to 3D to obtain relative pose/scale examples; (ii) it trains a text-conditioned score-based diffusion over the spatial parameter space and samples configurations at test time. As no implementation or checkpoints are available, we include this baseline only in the qualitative comparison (\cref{sec:qualitative}).

\textbf{SceneMotifCoder (SMC)}~\cite{tam2024smc}.
SMC learns meta-programs for 3D arrangements from a few examples; at inference it uses an LLM to instantiate program arguments, retrieves meshes, and runs a geometry-aware optimization to produce physically plausible placements.

\textbf{SceneTeller}~\cite{ocal2024sceneteller}.
SceneTeller is an LLM-driven scene arrangement method. Although designed for indoor scene layouts, its language-to-layout interface allows us to pose two-object cases as minimal ``scenes'', enabling a comparison on OOA scenarios.

\textbf{(B1) Shrinkwrap (single start).}
The source mesh is projected onto the target mesh using a shrinkwrap constraint~\cite{blenderShrinkwrap}, which attaches vertices to the nearest points on the target surface. 
This baseline is geometry-only and uses no text guidance.

\textbf{(B2) Shrinkwrap (multi-start + CLIP selection).}
The shrinkwrap procedure is repeated with $N=5$ randomized initial poses. Each resulting placement is scored using the CLIP similarity metric, and the best-scoring pose is selected. This variant retains a geometric projection step but incorporates semantic selection via CLIP, yielding a stronger baseline than B1.

\subsection{Quantitative Results}
\begin{figure}[t]
  \centering
  \begin{subfigure}{0.48\columnwidth}
    \includegraphics[width=\linewidth]{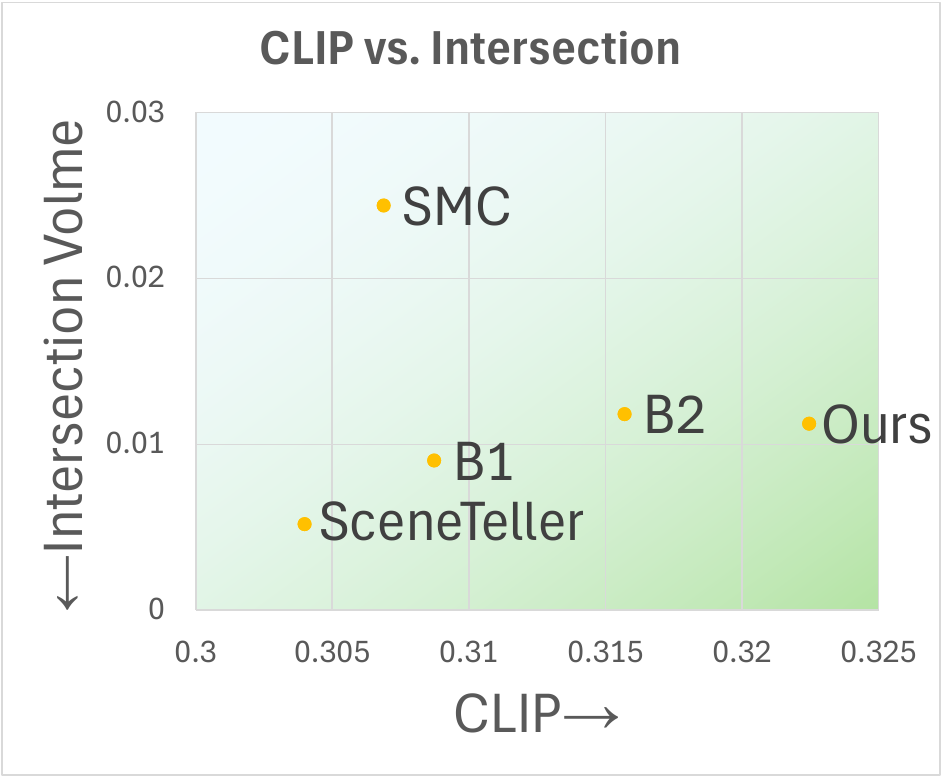}
  \end{subfigure}
  \begin{subfigure}{0.48\columnwidth}
    \includegraphics[width=\linewidth]{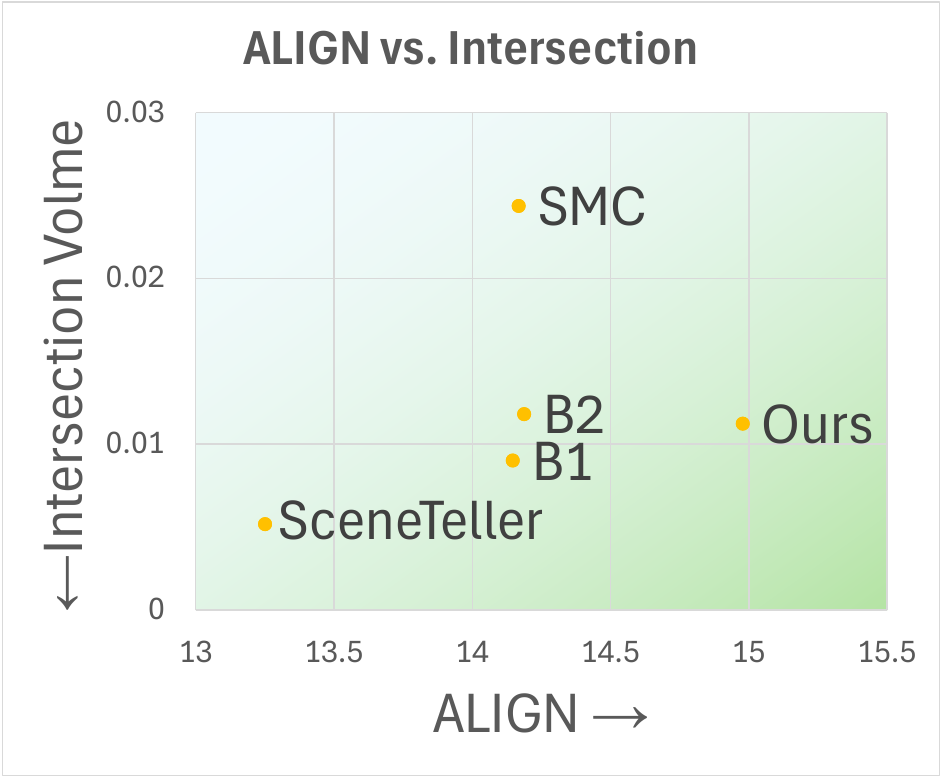}
  \end{subfigure}
  \caption{\textbf{Trade-off plot.} CLIP, ALIGN score vs.\ intersection volume score. Down and to the right is better. The full visualization, including SigLIP, is provided in the supplementary.}
  \label{fig:tradeoff}
\end{figure}

\cref{tab:quantitative} summarizes averages across the benchmark using the metrics in \cref{sec:metrics}. 
Our method attains the highest semantic scores on all three measures while maintaining competitive intersection volume. 
Notably, across the baselines, SceneTeller achieves the lowest intersection on average, but this often coincides with weaker semantic alignment, see the qualitative comparisons in \cref{fig:qualitative}, where placements that minimize penetration can still miss the intended interaction. 
As a trade-off summary, \cref{fig:tradeoff} plots CLIP and ALIGN versus intersection volume; in all panels our method lies bottom-right, indicating the best combination of high semantic agreement and low interpenetration. 
On the VLM-based evaluator, our method also ranks first among the baselines on all reported criteria: Text-Asset Alignment, 3D Plausibility, Text-Geometry Alignment, and Overall. 
We additionally report a scale-enabled variant, which is listed separately because competing baselines operate in a rigid setting and do not optimize scale.

\begin{table*}[t]
\centering
\setlength{\tabcolsep}{4pt}
\scriptsize
\begin{adjustbox}{max width=\textwidth}
\begin{tabular}{lcccccccc}
\toprule
\textbf{Method} & \textbf{CLIP} $\uparrow$ & \textbf{ALIGN} $\uparrow$ & \textbf{SigLIP} $\uparrow$ & \textbf{Intersection Volume} $\downarrow$ & \textbf{Text-Asset Alignment} $\uparrow$ & \textbf{3D Plausibility} $\uparrow$ & \textbf{Text-Geometry Alignment} $\uparrow$ & \textbf{Overall} $\uparrow$ \\
\midrule
Ours         & \textbf{0.3224} & 14.9800 & \textbf{0.0380} & 0.0112 & \textbf{1028.72} & \textbf{1024.06} & \textbf{1068.47} & \textbf{1034.44} \\
B1           & 0.3087 & 14.1484 & 0.0374 & 0.0090 & 988.63 & 1022.38 & 960.55 & 1005.92 \\
B2           & 0.3157 & 14.1897 & 0.0362 & 0.0118 & 997.41 & 987.63 & 980.59 & 1019.69 \\
SceneTeller  & 0.3040 & 13.2522 & 0.0367 & \textbf{0.0051} & 986.40 & 989.41 & 938.30 & 963.80 \\
SMC          & 0.3069 & 14.1704 & 0.0370 & 0.0244 & 985.75 & 991.30 & 937.56 & 991.19 \\
\midrule
Ours (+ scale)   & 0.3176 & \textbf{15.0350} & \textbf{0.0380} & 0.0110 & 994.52 & 959.33 & 1029.71 & 1005.10 \\
\bottomrule
\end{tabular}
\end{adjustbox}
\caption{\textbf{Baseline comparison.} Semantic alignment (CLIP/ALIGN/SigLIP; higher is better), physical plausibility (intersection volume; lower is better), and VLM evaluator scores (Text--Asset, 3D Plausibility, Text--Geometry, Overall; higher is better).}
\label{tab:quantitative}
\end{table*}

\subsection{Ablation Study}
\label{sec:ablation}
\begin{figure}[t]
  \centering
  \includegraphics[width=\columnwidth]{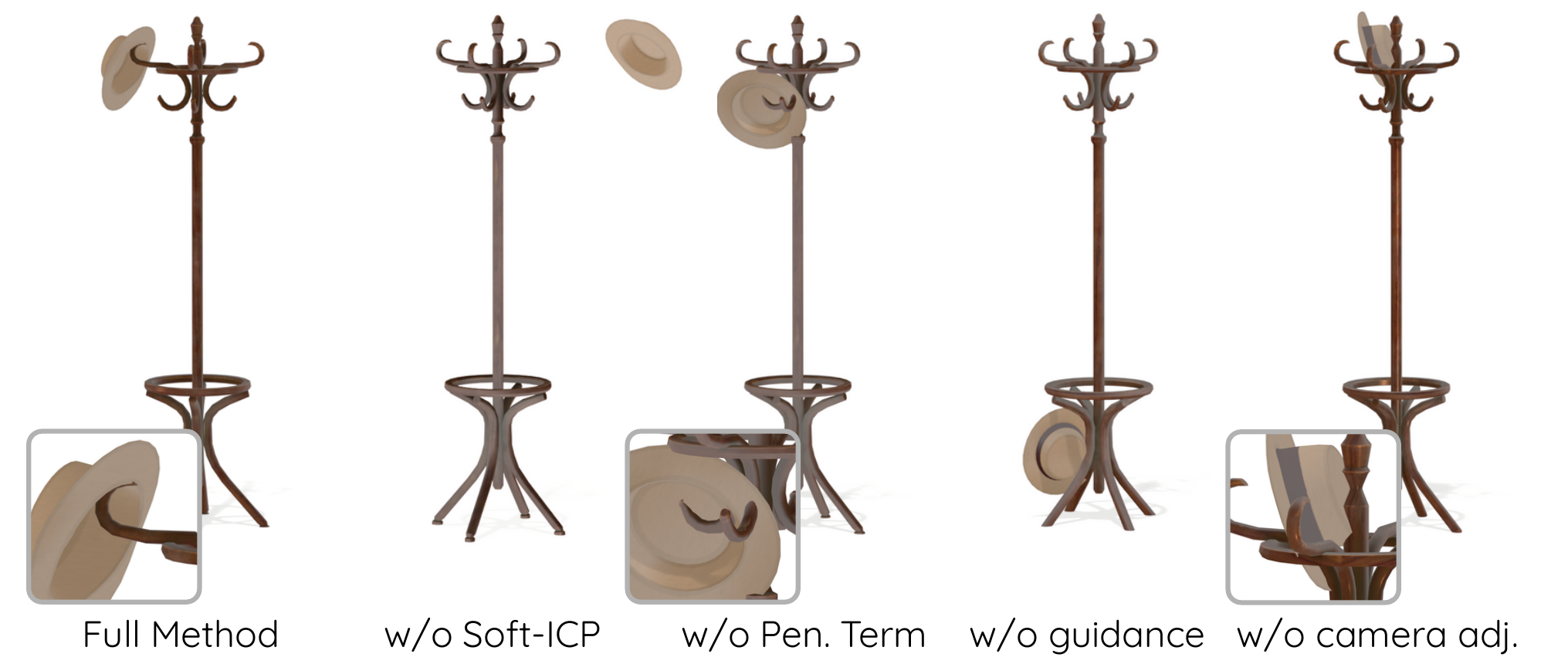}
  \caption{\textbf{Ablations on \emph{coatrack} (target) and \emph{hat} (source).}
  The full method yields the most plausible, text-aligned placement; ablating individual components produces degradations.
  }
  \label{fig:ablations_hanger_hat}
\end{figure}

We evaluate \textit{w/o guidance} (no semantic text guidance), \textit{w/o soft-ICP}, \textit{w/o penetration}, \textit{w/o phases}, \textit{w/o camera adjustment}, \textit{SDS} (replace CLIP with score distillation sampling~\cite{poole2022dreamfusion}), and \textit{SigLIP-guidance} (replace CLIP with SigLIP). Quantitative comparison appear in \cref{tab:quantitative_ablations}.
The \textit{w/o camera adjustment} variant appears close on averages because its benefit is most pronounced for large size-ratio pairs which is a minority of the benchmark. 
We hypothesize SDS underperforms in this setting because its stochastic, synthesis-oriented gradients provide weaker directional signals for pose updates compared to CLIP’s contrastive image-text similarity.
\cref{fig:ablations_hanger_hat} illustrates these effects on a \emph{coatrack \& hat} example: the full model yields the most physically plausible, text-aligned placement, while ablations exhibit the characteristic failures described above.

\begin{table*}[t]
\centering
\setlength{\tabcolsep}{4pt}
\scriptsize
\begin{adjustbox}{max width=\textwidth}
\begin{tabular}{lcccccccc}
\toprule
\textbf{Method} & \textbf{CLIP} $\uparrow$ & \textbf{ALIGN} $\uparrow$ & \textbf{SigLIP} $\uparrow$ & \textbf{Intersection Volume} $\downarrow$ & \textbf{Text-Asset Alignment} $\uparrow$ & \textbf{3D Plausibility} $\uparrow$ & \textbf{Text-Geometry Alignment} $\uparrow$ & \textbf{Overall} $\uparrow$ \\
\midrule
Ours         & 0.3224 & 14.9800 & 0.0380 & 0.0112 & 1028.72 & 1024.06 & 1068.47 & 1034.44 \\
w/o guidance     & 0.3091 & 13.8372 & 0.0372 & 0.0099 & 985.52 & 979.31 & 959.38 & 999.79 \\
w/o soft-ICP     & 0.3214 & 14.7183 & 0.0368 & 0.0002 & 969.40 & 1031.15 & 976.74 & 1004.89 \\
w/o penet.       & 0.3223 & 14.7342 & 0.0376 & 0.0177 & 1007.71 & 1000.49 & 961.71 & 987.69 \\
w/o phases       & 0.3214 & 14.7627 & 0.0379 & 0.0141 & 978.24 & 965.52 & 974.38 & 996.56 \\
w/o camera adj.  & 0.3201 & 14.5859 & 0.0368 & 0.0127 & 1035.74 & 1019.02 & 1004.62 & 1023.30 \\
SDS              & 0.3167 & 14.4120 & 0.0377 & 0.0148 & 1022.78 & 1002.96 & 1010.46 & 1005.47 \\
SigLIP           & 0.3145 & 14.4185 & 0.0387 & 0.0136 & 1015.86 & 1014.75 & 1019.08 & 1025.52 \\
\bottomrule
\end{tabular}
\end{adjustbox}
\caption{\textbf{Ablation comparison.} Ablated variants across semantic alignment (CLIP/ALIGN/SigLIP; higher is better), physical plausibility (intersection volume; lower is better), and VLM evaluator scores (Text--Asset, 3D Plausibility, Text--Geometry, Overall; higher is better).}
\label{tab:quantitative_ablations}
\end{table*}

\subsection{Qualitative Results}\label{sec:qualitative}
\begin{figure}[t]
  \centering
  \includegraphics[width=\columnwidth]{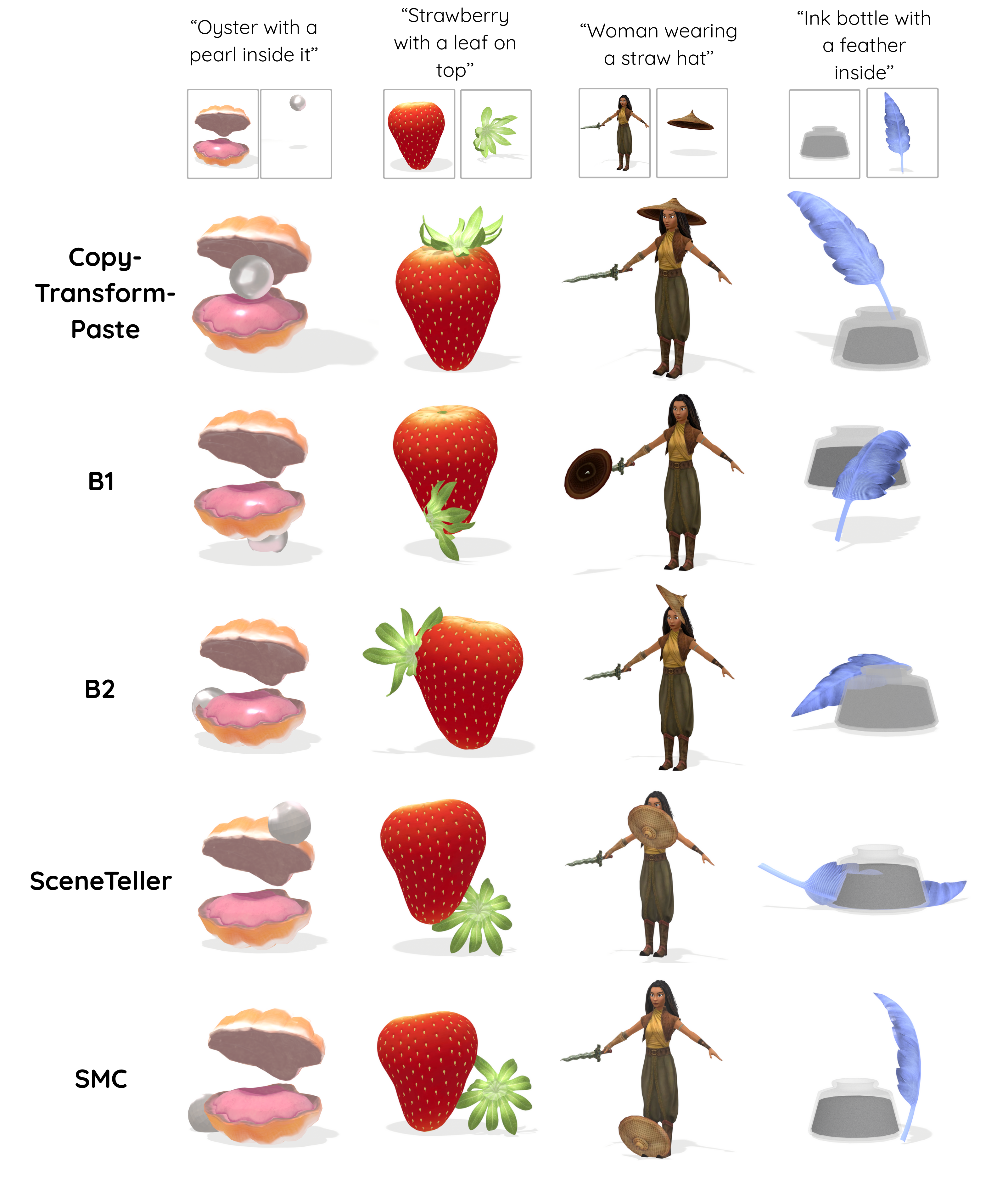}
  \caption{\textbf{Qualitative comparison across four object--object pairs.}
  \textit{Top row:} input prompt and meshes. 
  \textit{Rows 1–5:} final placements from our method and baselines. 
  Our approach yields semantically faithful and physically plausible alignments, while baselines vary in semantic faithfulness and contact quality.}
  \label{fig:qualitative}
\end{figure}

\cref{fig:qualitative} compares our method with \emph{B1}, \emph{B2}, \emph{SceneTeller}, and \emph{SMC}. 
Our method consistently produces semantically faithful, physically plausible placements. B1, SceneTeller and SMC often misplace the source on irrelevant regions or exhibit interpenetrations, while B2 improves semantic consistency but still shows physical or semantic mismatches.

\begin{figure}[t]
  \centering
  \includegraphics[width=\columnwidth]{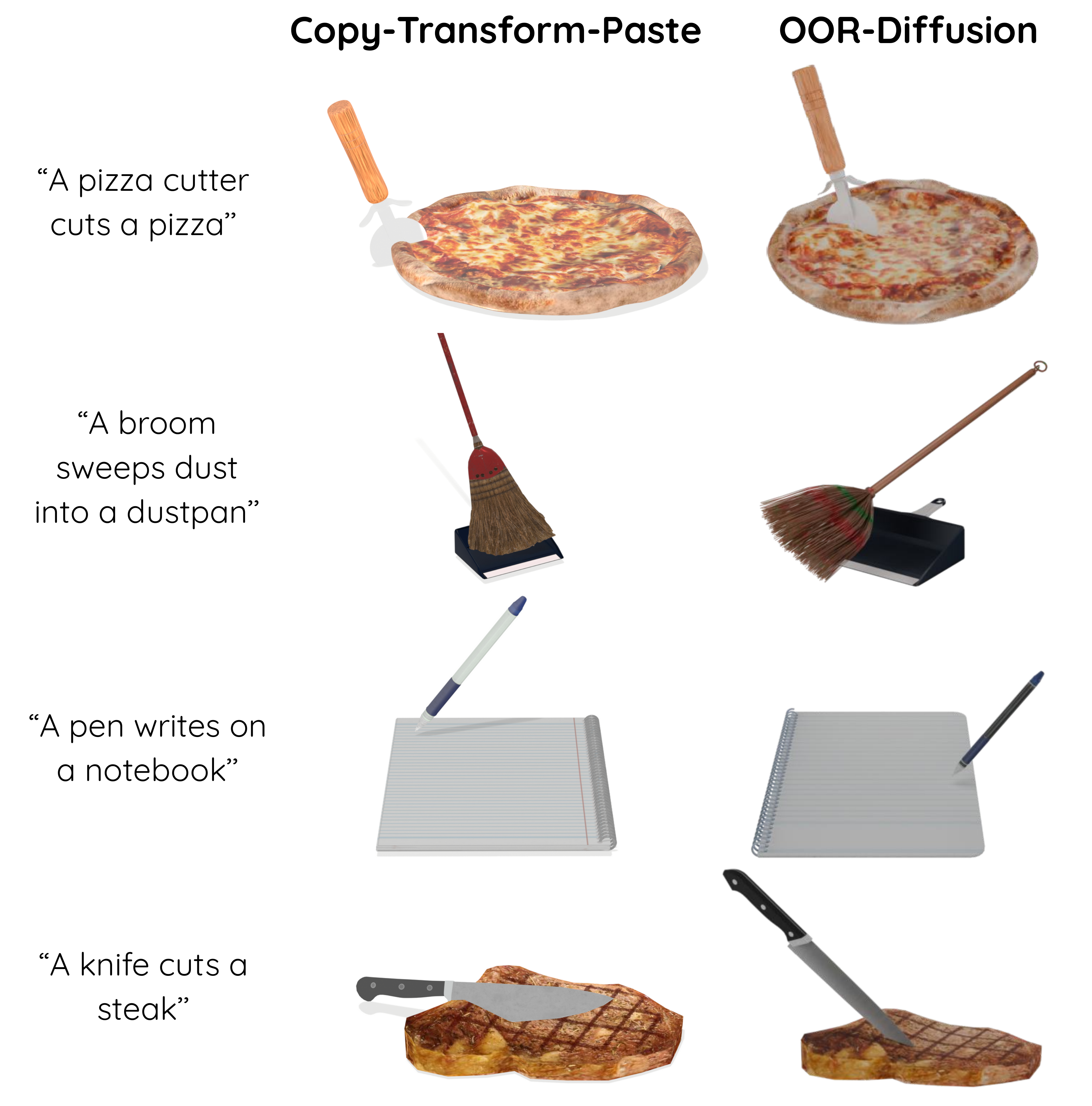}
  \caption{\textbf{Qualitative comparison to OOR-diffusion.}
  OOR-diffusion panels are reproduced from their paper; we matched assets and camera setups where possible. 
  }
  \label{fig:qualitative_oor}
\end{figure}

We also include a side-by-side with \emph{OOR-diffusion} (\cref{fig:qualitative_oor}). 
As no implementation or checkpoints are available, we extracted \emph{OOR-diffusion} images from their paper and qualitatively compare on those highlighted examples. 
For these cases, both methods produce text-aligned and physically plausible results. 
Where possible, we reproduced our results with similar assets and camera poses, though exact assets/cameras may differ.


\subsection{User study}
We conducted a user study comparing our method to four baselines (B1, B2, SceneTeller, SMC) on 15 object-pair instances, randomly selected from our benchmark, with $47$ participants. In each trial, participants were shown (i) the names of the two objects, (ii) the text prompt describing their interaction, and (iii) five rendered images, one per method, presented together. For every trial they answered two multiple-choice questions: (1) \emph{Which image best matches the textual description of the interaction?} and (2) \emph{In which image the two objects appear most physically plausible?} Method order and trial order were randomized per participant. Results appear in \cref{tab:userstudy}. 
Full protocol details appear in the Supplementary.

\begin{table}[t]
\centering
\setlength{\tabcolsep}{5pt}
\small
\resizebox{\columnwidth}{!}{%
\begin{tabular}{lccccc}
\toprule
\textbf{Question} & \textbf{Ours} & \textbf{B1} & \textbf{B2} & \textbf{SceneTeller} & \textbf{SMC} \\
\midrule
Matches description (\%) & \textbf{85.24} & 3.32 & 7.98 & 1.73 & 1.73 \\
Physically plausible (\%)& \textbf{79.65} & 3.86 & 8.78 & 1.99 & 5.72 \\
\bottomrule
\end{tabular}
}
\caption{\textbf{User study results.} Percentage of votes across $15$ instances and $47$ participants. Higher is better.}
\label{tab:userstudy}
\end{table}

\subsection{Evaluation of LLM hyperparameter selection}
\label{sec:llm_eval}
We evaluate the LLM on a held-out set of 61 object--pair instances with ground-truth (GT) labels equal to the hyper-parameters defined in \cref{subsec:hparam_llm}: (i) initial scale ratio, (ii) penetration policy, and (iii) attachment ratio. For each instance, the LLM receives only the two object names and the text prompt and returns one value per hyperparameter. We repeated 10 independent queries per instance. We report accuracy for the binary penetration label, and Mean Absolute Error (MAE) for the continuous quantities: initial scale and attachment ratio (\cref{tab:llm_eval}).

\begin{table}[t]
\centering
\small
\setlength{\tabcolsep}{6pt}
\resizebox{\columnwidth}{!}{%
\begin{tabular}{lccc}
\toprule
& \textbf{Penet.\ (Acc.\%\,$\uparrow$)} & \textbf{Init.\ scale (MAE$\downarrow$)} & \textbf{Attach.\ ratio (MAE$\downarrow$)} \\
& \textit{boolean} & \textit{$[0.1,10]$} & \textit{$[0,1]$} \\
\midrule
\textbf{LLM (ours)} & 98.36 & 0.46 & 0.25 \\
\textbf{Random}     & 50.00 & 3.30 & 0.33 \\
\bottomrule
\end{tabular}
}
\caption{\textbf{LLM hyperparameter selection.}
Results on 61 held-out object--pair instances.
The LLM predicts three hyperparameters: penetration policy, initial scale ratio, and attachment ratio.The random baseline corresponds to expectations from uniform sampling over the indicated ranges.}
\label{tab:llm_eval}
\end{table}

\section{Other applications}
\label{sec:applications}

\begin{figure}[t]
  \centering
  \includegraphics[width=\columnwidth]{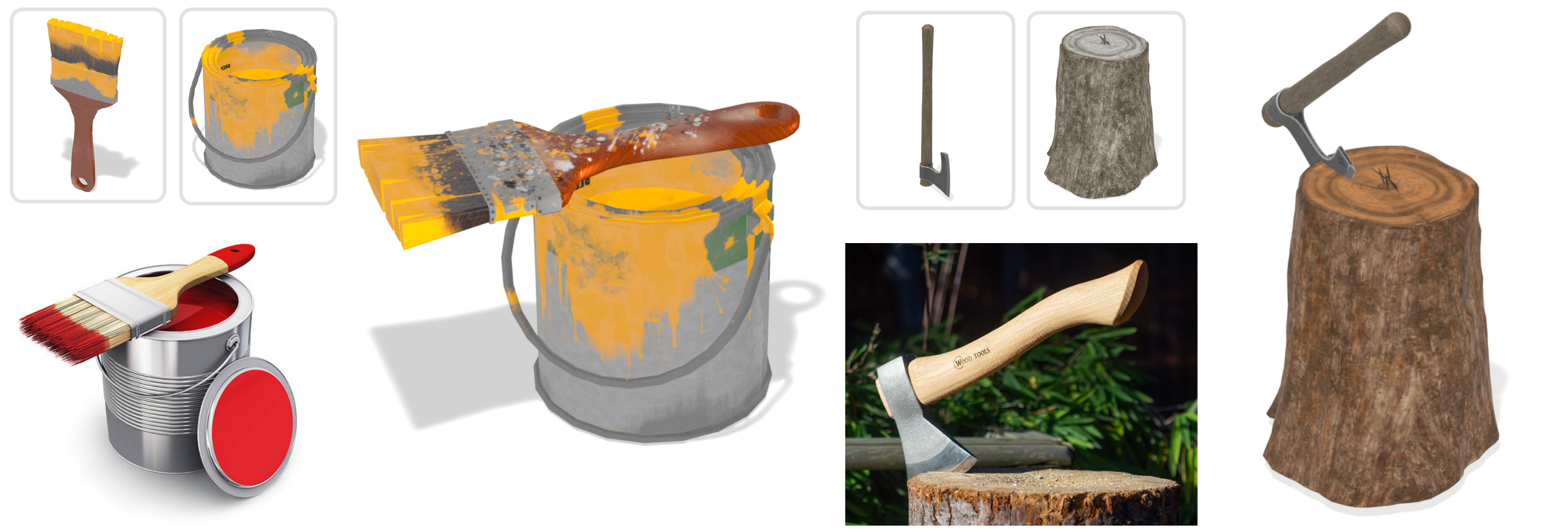}
  \caption{\textbf{Image-to-3D alignment.} Inputs: image + two meshes. Output: aligned meshes whose renders match the image.
  }
  \label{fig:app_image2mesh}
\end{figure}

\noindent\textbf{Image-to-3D alignment.}
Given an image and two meshes, we render candidate views and use the same language-vision objectives to align the meshes so that the rendered views agree with the image description (see \cref{fig:app_image2mesh}).

\noindent\textbf{Multi-object assembly (iterative).}
We can align multiple objects by iterating: the output of stage \(k\) becomes part of the input at stage \(k{+}1\), progressively assembling the scene (see \cref{fig:teaser}, bottom-right).


\section{Limitations}
\label{sec:limitations}
While the proposed method demonstrates robust and semantically coherent object-object alignment, several limitations remain.

\textbf{Penetration residuals.}
Despite the penetration loss, the final placements may exhibit minor interpenetrations. Increasing the penetration weight in later phases or re-running with different random initializations often mitigates this.

\textbf{View scheduling and finite-view supervision.}
Because cameras are re-targeted across phases (\cref{subsec:camera}) and gradients are computed from a finite set of views, viewpoint-sensitive predicates such as \emph{“beside/next to/left of/right of”} can be unstable: the lateral frame of reference shifts with the current view, and solutions that look correct from the sampled angles may still be wrong in 3D.

\textbf{Extreme scale differences and partial visibility.}
When objects differ greatly in size, the smaller one may occupy a negligible fraction of the image, yielding unreliable language-vision gradients even with zoom-in scheduling.
Similar degradation appears under heavy occlusion (\eg, insertion into a cavity). These effects align with reported CLIP biases toward the more salient or larger object in multi-object scenes~\cite{Abbasi_2025_CVPR}.

\begin{figure}[t]
  \centering
  \includegraphics[width=\columnwidth]{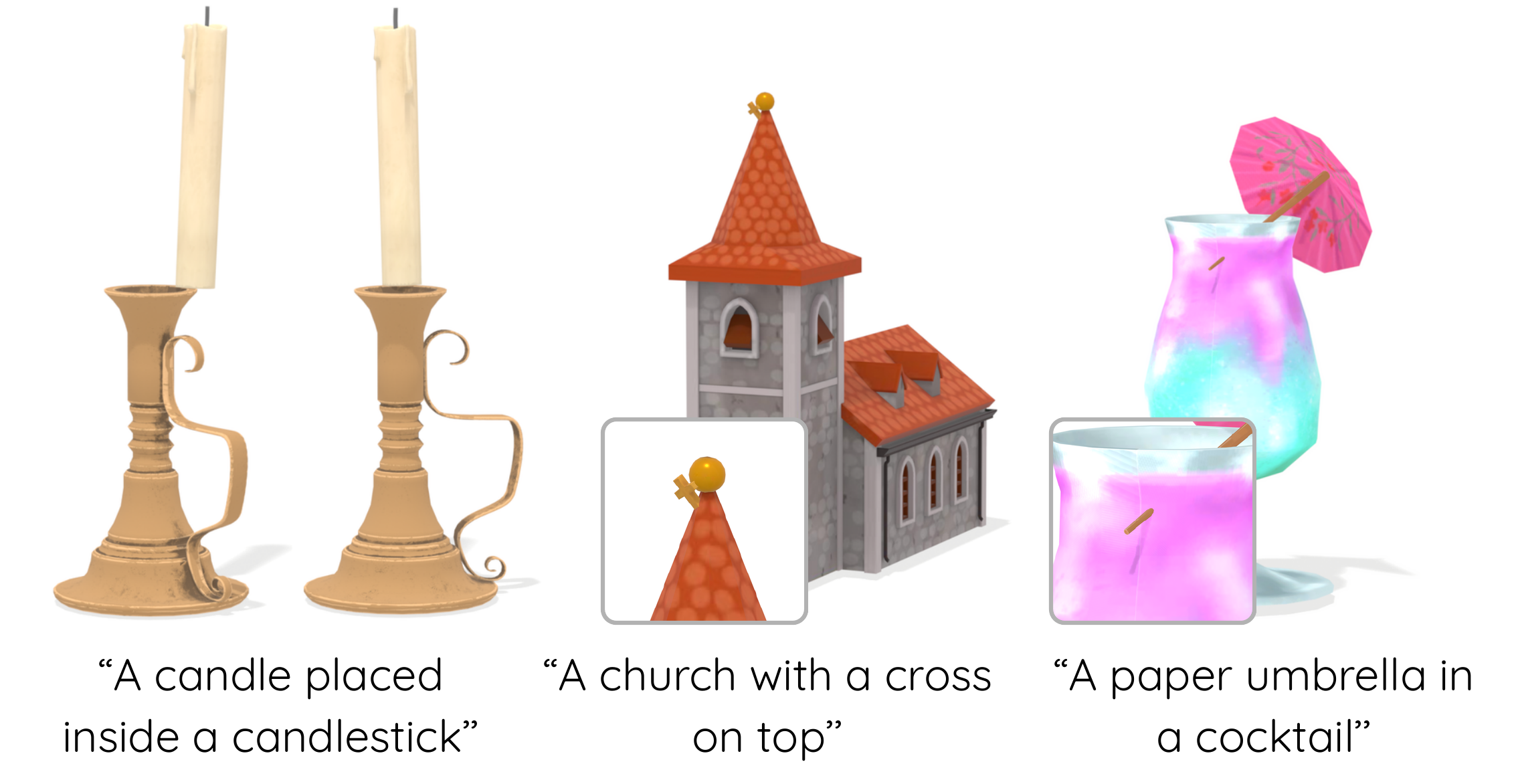}
  \caption{\textbf{Failure cases.} Errors from multi-view ambiguity and insertion into cavity, extreme scale mismatch that weakens language guidance, and residual interpenetration (often mitigated by re-running or increasing the penetration weight).
  }
  \label{fig:failures}
\end{figure}

\cref{fig:failures} illustrates representative cases for viewpoint sensitivity, extreme scale, and penetration residuals.

\section{Conclusion}
This work introduced a text-guided framework for aligning two 3D meshes under vision-language supervision and geometric constraints.  
By optimizing translation, rotation, and scale through differentiable rendering, the method achieves semantically faithful and physically plausible arrangements.  
A benchmark of text-mesh-mesh triplets enables standardized evaluation of text-guided object alignment.  
Extensive experiments demonstrate robustness and highlight the roles of geometric and language-based objectives. Future work will explore stronger vision-language models, multi-view consistency, and physics-based reasoning to further improve realism.

{
    \small
    \bibliographystyle{ieeenat_fullname}
    \bibliography{main}
}

\clearpage
\setcounter{page}{1}
\maketitlesupplementary

\section{Benchmark Construction}
\label{sec:supp_benchmark}

\paragraph{Two curated benchmarks.}
We release two OOA variants: \textbf{(i) Rigid} (translation+rotation only; no scale) and \textbf{(ii) Scale-enabled} (translation+rotation+isotropic scale). This separation enables fair comparison to methods that do not support scale.

\paragraph{Source assets and reference pose.}
We collected pairs of meshes from Sketchfab. For each pair, we manually placed the objects in their intended relational configuration (e.g., candle inside candlestick). This serves as the reference pose. It fixes only the relative arrangement; meshes are not necessarily canonicalized or upright.

\paragraph{Perturbation protocol (initializations).}
Starting from the reference pose, we create randomized starting positions by selecting one object at random and:
\begin{itemize}[leftmargin=*,nosep]
  \item \textbf{Translate} each mesh independently by a random offset 
\(\Delta \sim \mathcal{U}\!\big(-10\,\mathbf{L},\, 10\,\mathbf{L}\big)\),
where \(\mathbf{L}=(L_x,L_y,L_z)\) are the side lengths of its axis-aligned bounding box.
  \item \textbf{Rotate} it by independent Euler angles within \(\pm 180^\circ\).
  \item \textbf{Scale} (for the \emph{scale-enabled} benchmark only) isotropically, sampled uniformly from \([0.01,\,100.0]\).
\end{itemize}
All draws are uniform and independent; the \emph{rigid} benchmark keeps the scale fixed at \(1.0\).

\paragraph{Outputs.}
For each benchmark instance we also provide: (1) the \emph{final aligned meshes}, and (2) the \emph{text prompt}.

\section{LLM prompt design for hyperparameter selection}
\label{supsec:llm_prompts}

In \cref{subsec:hparam_llm}, we described how an LLM is used to estimate three hyperparameters from the object names and scene description.  
Here, we provide the exact prompts used to query the model.  
Each prompt is written in natural language and instructs the model to output a single JSON value on one line.

\paragraph{(1) Initial scale.}
Estimates the real-world relative size between the two objects:
\begingroup
\footnotesize
\ttfamily
\noindent Estimate the relative scale needed so that object1="{\{}object1{\}}" and object2="{\{}object2{\}}" fit together naturally in the desired alignment "{\{}wanted\_alignment{\}}".  
\noindent Define size\_ratio = bbox\_size(object1) / bbox\_size(object2).  
\noindent Output exactly one JSON object: \{"size\_ratio": <float between 0.01 and 100.0>\}.
\endgroup

\paragraph{(2) Penetration policy.}
Determines whether the final configuration should involve one object penetrating another (e.g., cutting, slicing):
\begingroup
\footnotesize
\ttfamily
\noindent Decide whether achieving alignment "{\{}wanted\_alignment{\}}" between object1="{\{}object1{\}}" and object2="{\{}object2{\}} REQUIRES solid-to-solid penetration.  
\noindent Output exactly one JSON object: \{"penetration": <true|false>\}.
\endgroup

\paragraph{(3) Attachment ratio.}
Estimates how much surface contact is expected between the two objects in the final arrangement:
\begingroup
\ttfamily\footnotesize
\noindent For the desired alignment "{\{}wanted\_alignment{\}}", estimate the fraction of surface contact between object1="{\{}object1{\}}" and object2="{\{}object2{\}}".
\noindent Define contact\_ratio in [0,1], where 0 = almost no contact and 1 = full surface contact.
\noindent Output exactly one JSON object: \{"contact\_ratio": <float 0..1>\}.\par
\endgroup

\paragraph{Usage.}
At test time, we query the LLM with these three templates using only the object names and textual prompt, without any rendered images.  
The model outputs the three values used to set: (i) the source-object initial scale, (ii) whether the penetration loss is enabled, and (iii) the attachment ratio \(r\) controlling the soft-ICP vertex fraction.

\section{User Study Protocol}
\label{sec:supp_user_study}
\begin{figure*}[t]
  \centering
  \includegraphics[width=0.7\textwidth]{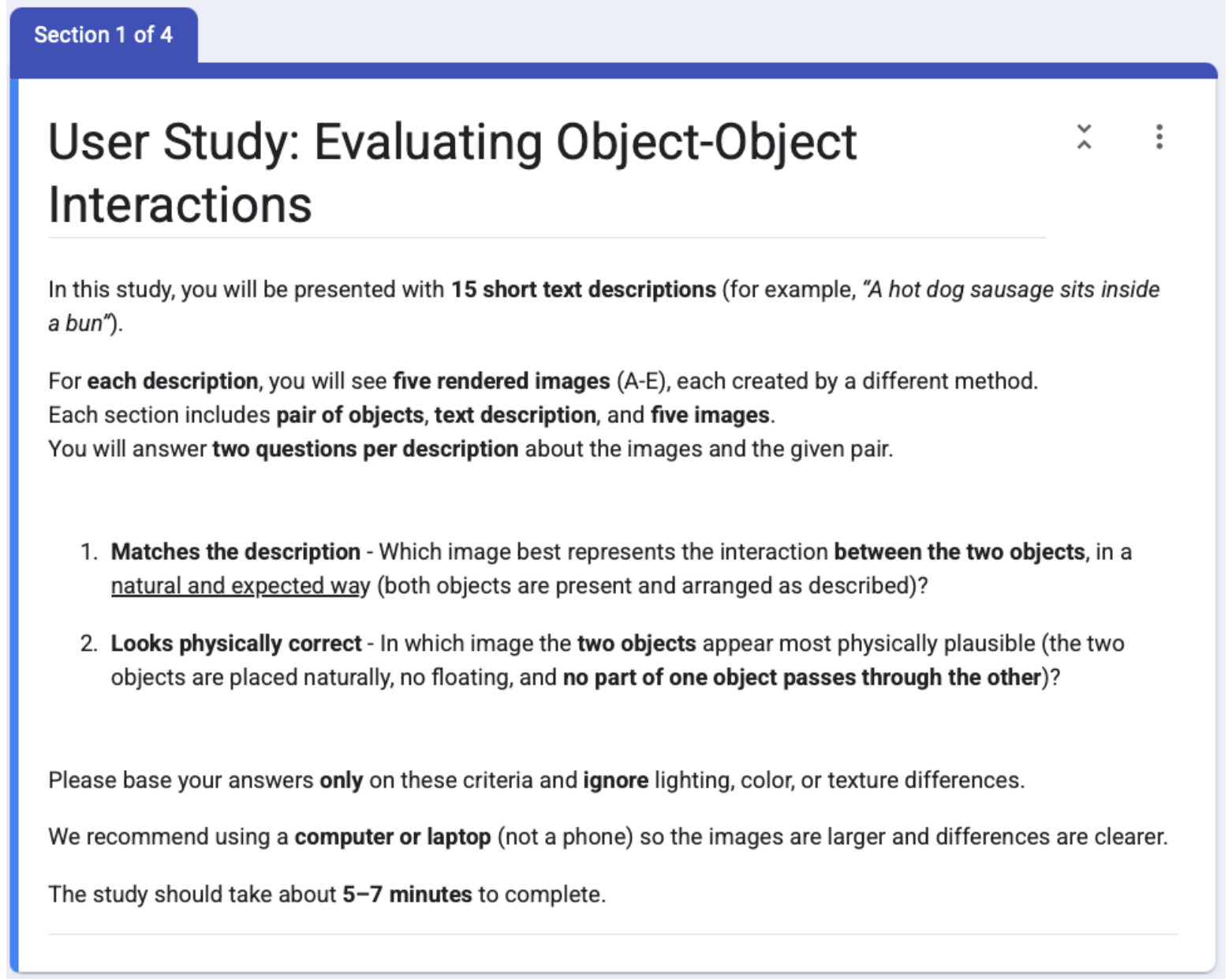}
  \caption{\textbf{User study guidelines (screenshot).}}
  \label{fig:user_study_guidelines}
\end{figure*}

\begin{figure*}[t]
  \centering
  \includegraphics[width=0.48\textwidth]{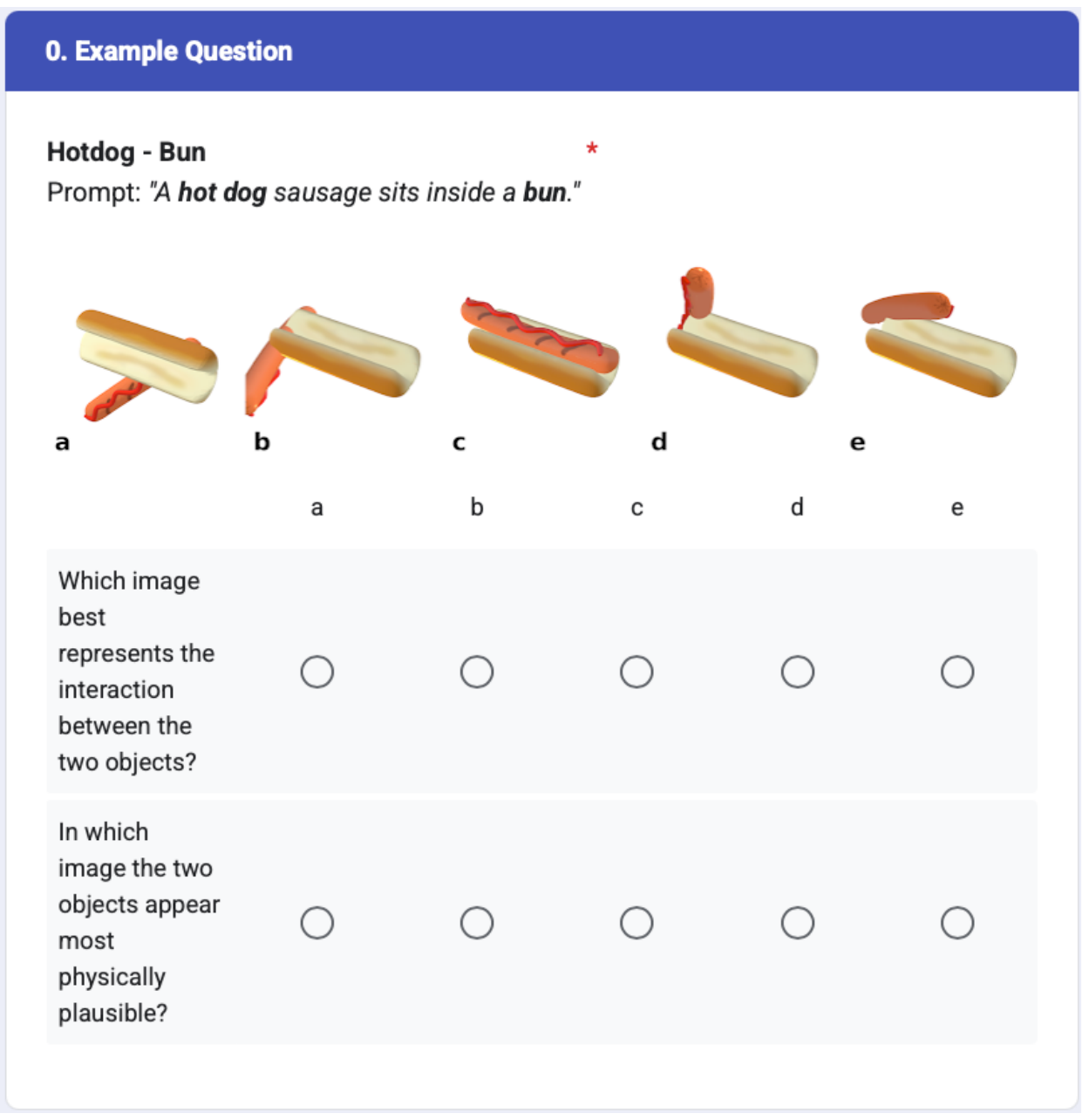}
  \vspace{-4em}
  \caption{\textbf{Example trial (screenshot).}}
  \label{fig:user_study_example}
\end{figure*}

We include here the full user study protocol used in our evaluation, consisting of the instructions shown to participants (Fig.~\ref{fig:user_study_guidelines}) and an example trial interface (Fig.~\ref{fig:user_study_example}).
These materials were presented exactly as shown to all participants during the study.

\section{Additional Trade-off Visualizations}
\label{sec:supp_tradeoff}

\begin{figure}[t]
  \centering
  \begin{subfigure}{0.48\linewidth}
    \includegraphics[width=\linewidth]{imgs/clip_vs_intersection_gradient.pdf}
    \caption{CLIP vs.\ intersection}
  \end{subfigure}
  \hfill
  \begin{subfigure}{0.48\linewidth}
    \includegraphics[width=\linewidth]{imgs/align_vs_intersection_gradient.pdf}
    \caption{ALIGN vs.\ intersection}
  \end{subfigure}

  \begin{subfigure}{0.48\linewidth}
    \includegraphics[width=\linewidth]{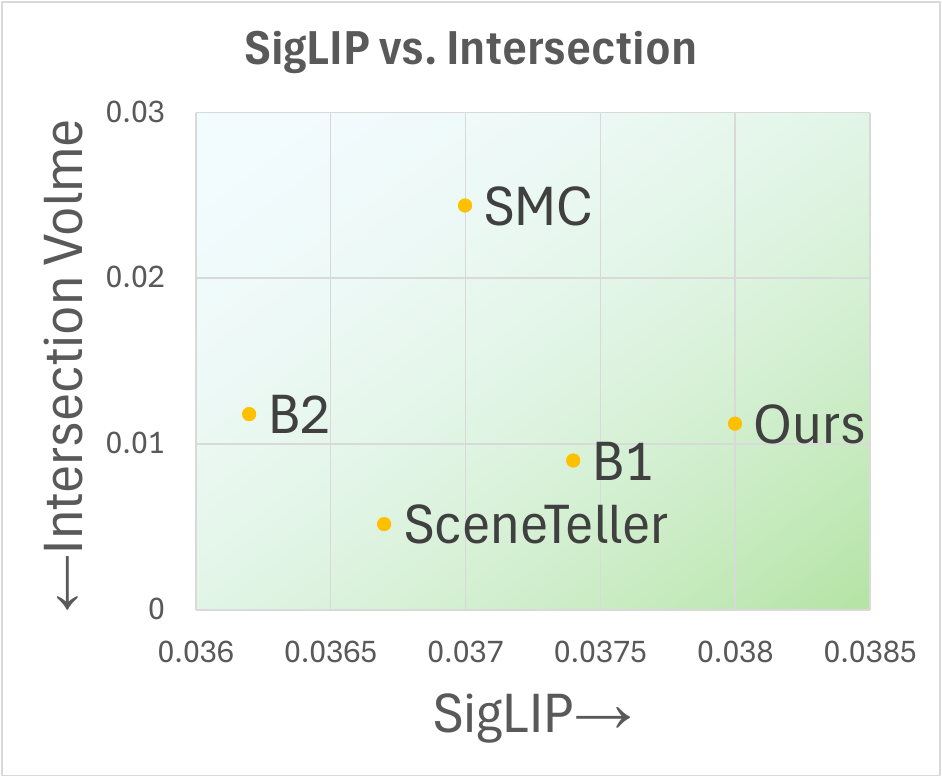}
    \caption{SigLIP vs.\ intersection}
  \end{subfigure}
  \caption{\textbf{Extended trade-off plots.}
  Vision–language alignment score versus intersection volume. Lower intersection and higher alignment (down-right) is better.}
  \label{fig:tradeoff_supp}
\end{figure}

Figure~\ref{fig:tradeoff_supp} provides the full trade-off visualization referenced in the main paper, including results for SigLIP in addition to CLIP and ALIGN.

\section{Additional Image-to-3D Alignment Results}
\label{sec:supp_image_to_3d}

\begin{figure*}[t]
  \centering
  \includegraphics[width=\textwidth]{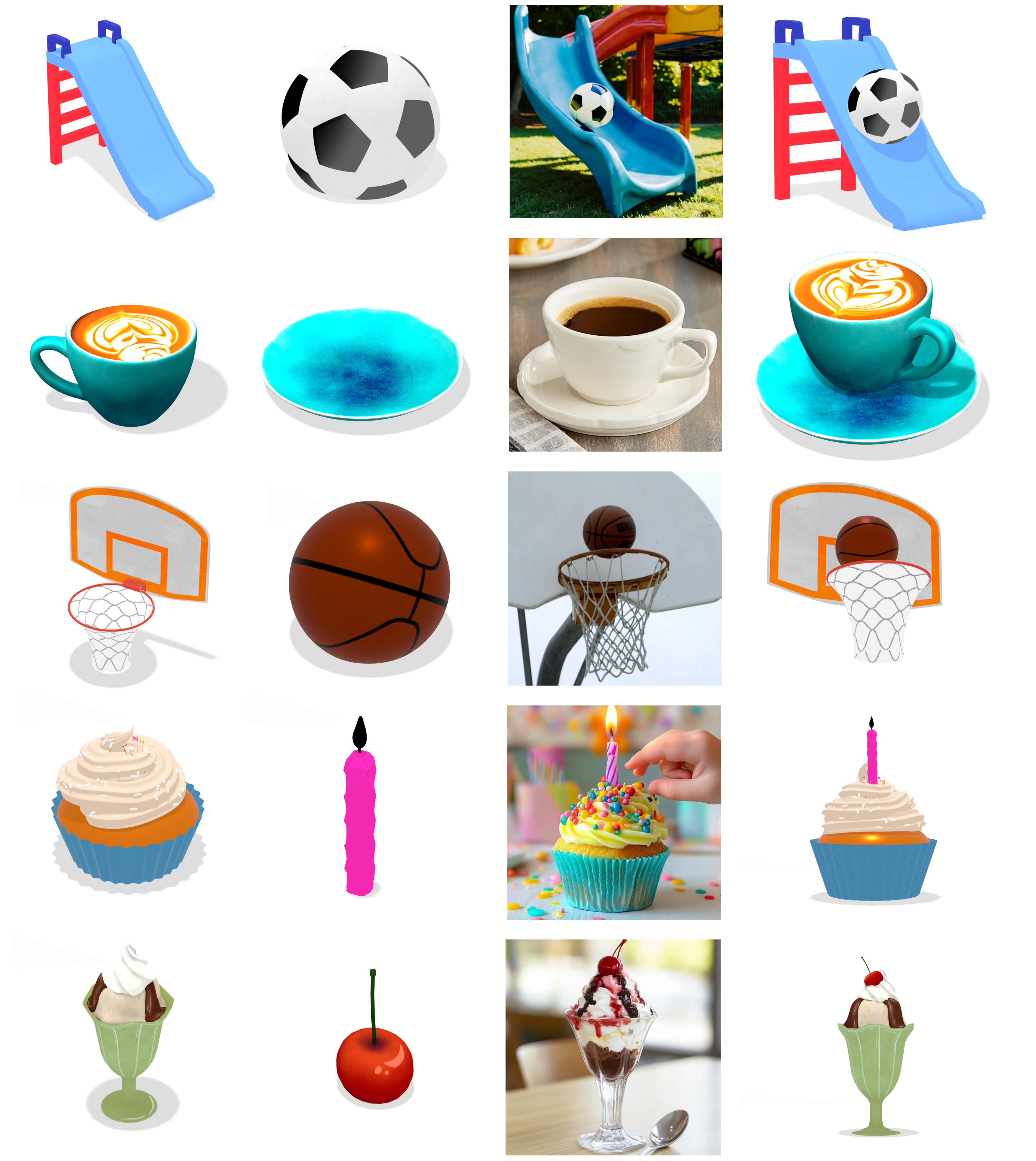}
  \caption{\textbf{Additional image-to-3D alignment results.}
  Each row shows an example of our image-guided alignment process: 
  the two input meshes (left), the reference image used for guidance (middle), 
  and the final optimized placement (right). 
  These examples illustrate the model’s ability to interpret image cues 
  and produce semantically consistent 3D arrangements.}
  \label{fig:image_to_3d_more}
\end{figure*}

Beyond text-conditioned alignment, our method can also be guided by reference images. 
Figure~\ref{fig:image_to_3d_more} presents additional examples of this image-to-3D setting.  
Each row shows the two input meshes, the guiding reference image, and the final optimized arrangement.
\section{Additional Qualitative Comparisons}
\label{sec:supp_more_qualitative}
\begin{figure*}[t]
  \centering
  {\large \textbf{``Ice cream sits inside a cone''}} \\[8pt]
  \begin{minipage}[t]{0.12\textwidth}
    \vspace{37pt} 
    \raggedleft
    \footnotesize
    \textbf{Ours} \\[74pt]
    \textbf{B1} \\[74pt]
    \textbf{B2} \\[74pt]
    \textbf{SceneTeller} \\[74pt]
    \textbf{SMC}
  \end{minipage}
  \hfill
  \begin{minipage}[t]{0.85\textwidth}
    \vspace{0pt}
    \includegraphics[width=\linewidth]{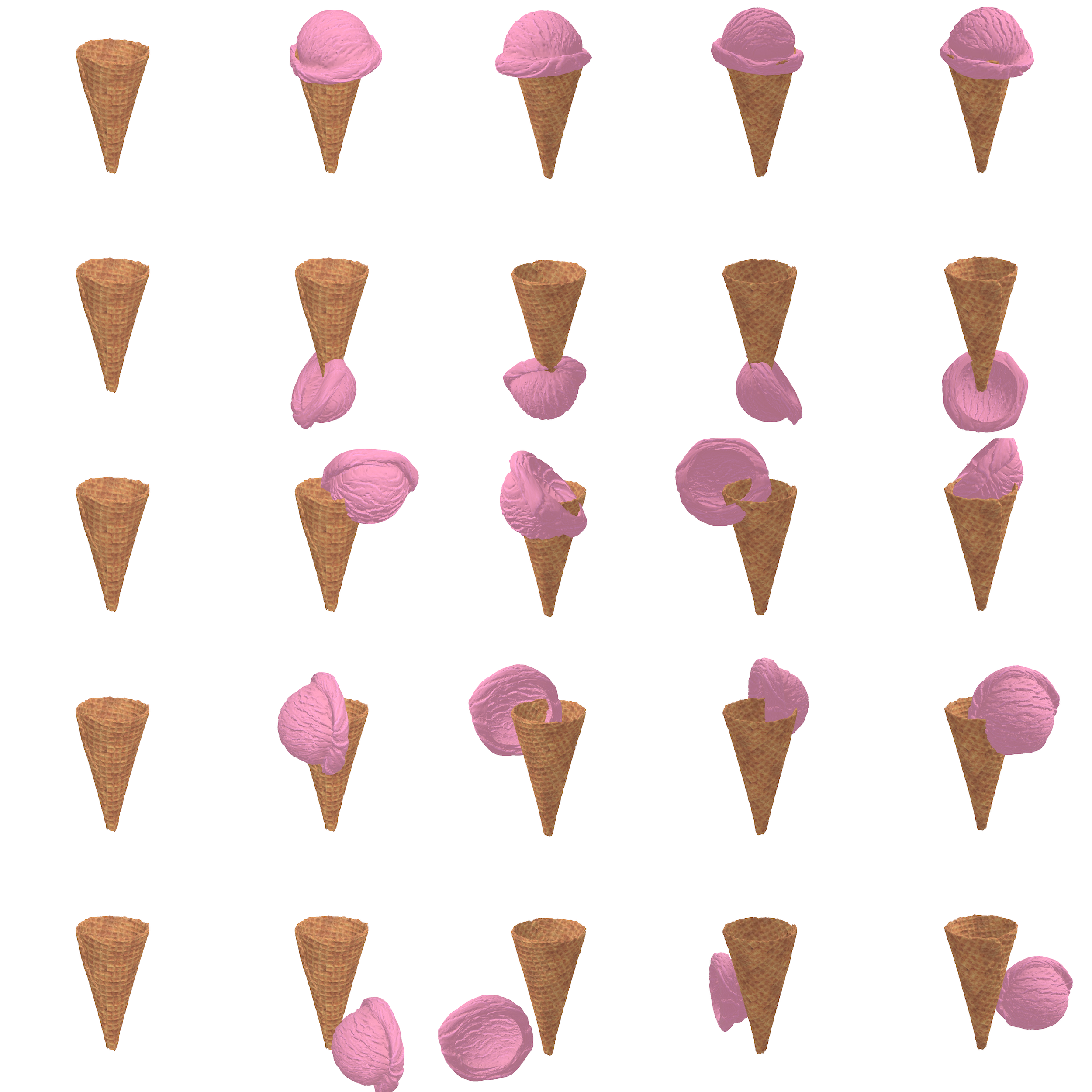}
  \end{minipage}
  \caption{\textbf{Per-mesh qualitative comparison: ice cream and cone.}
    Rows correspond to our method, B1, B2, SceneTeller, and SMC (top to bottom).
    In each row, the leftmost panel shows the target mesh alone, followed by the final aligned meshes rendered from four different viewing angles.
    Only our method achieves a text-aligned configuration where the ice cream sits physically inside the cone.}
  \label{fig:ice_cream_qual_supp}
\end{figure*}

\begin{figure*}[t]
  \centering
  {\large \textbf{``A blender jar sits on top of the blender base''}} \\[8pt]
  \begin{minipage}[t]{0.12\textwidth}
    \vspace{37pt} 
    \raggedleft
    \footnotesize
    \textbf{Ours} \\[74pt]
    \textbf{B1} \\[74pt]
    \textbf{B2} \\[74pt]
    \textbf{SceneTeller} \\[74pt]
    \textbf{SMC}
  \end{minipage}
  \hfill
  \begin{minipage}[t]{0.85\textwidth}
    \vspace{0pt}
    \includegraphics[width=\linewidth]{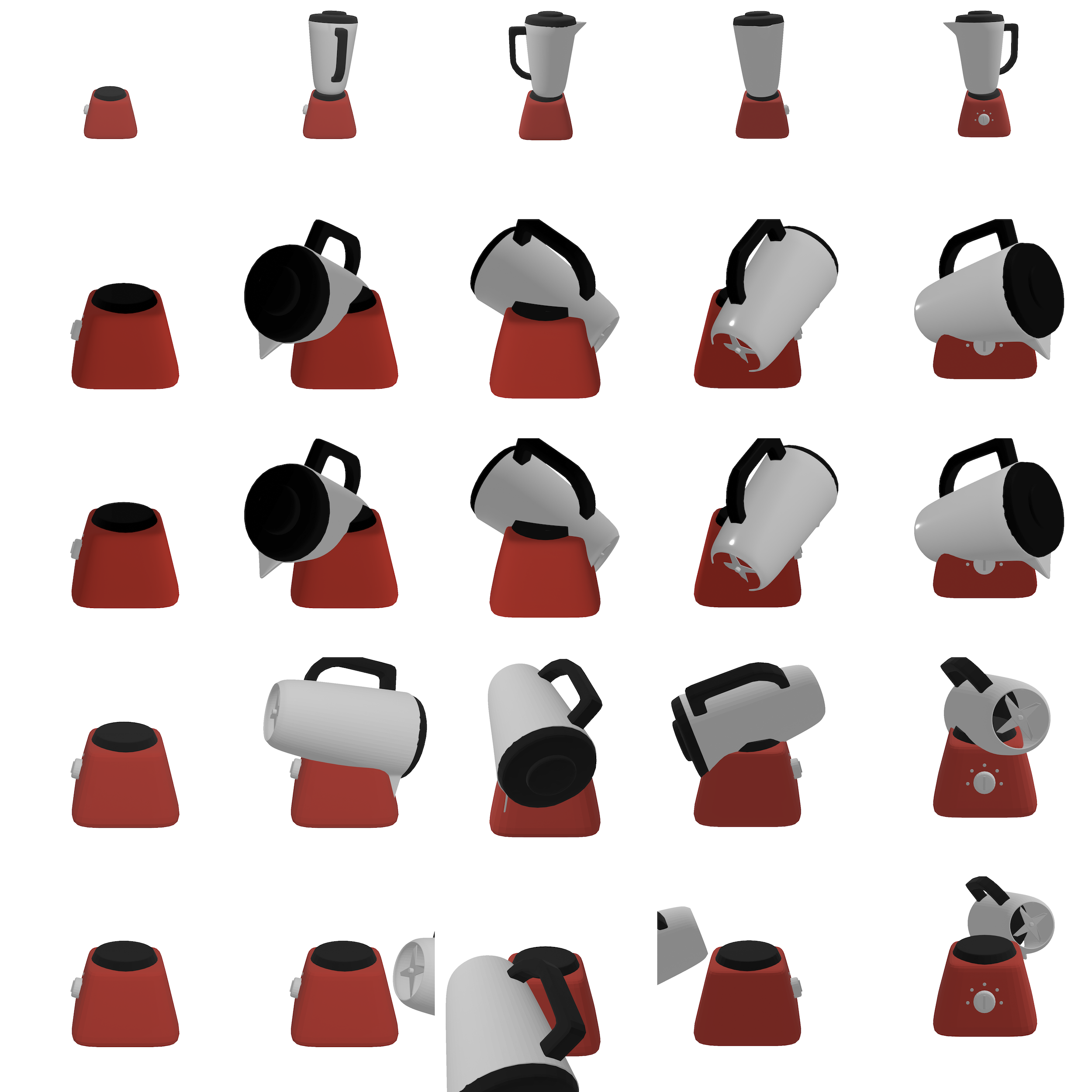}
  \end{minipage}
  \caption{\textbf{Per-mesh qualitative comparison: blender jar and base.}
Rows correspond to our method, B1, B2, SceneTeller, and SMC (top to bottom).
In each row, the leftmost panel shows the target mesh alone, followed by the final aligned meshes rendered from four different viewing angles.
Only our method places the jar upright and stably on the base; other baselines do put the jar on top but typically tilt it and exhibit noticeable interpenetrations between jar and base.}
  \label{fig:blender_qual_supp}
\end{figure*}

\begin{figure*}[t]
  \centering
  {\large \textbf{``Hot dog sausage sits inside a bun''}} \\[8pt]
  \begin{minipage}[t]{0.12\textwidth}
    \vspace{37pt} 
    \raggedleft
    \footnotesize
    \textbf{Ours} \\[74pt]
    \textbf{B1} \\[74pt]
    \textbf{B2} \\[74pt]
    \textbf{SceneTeller} \\[74pt]
    \textbf{SMC}
  \end{minipage}
  \hfill
  \begin{minipage}[t]{0.85\textwidth}
    \vspace{0pt}
    \includegraphics[width=\linewidth]{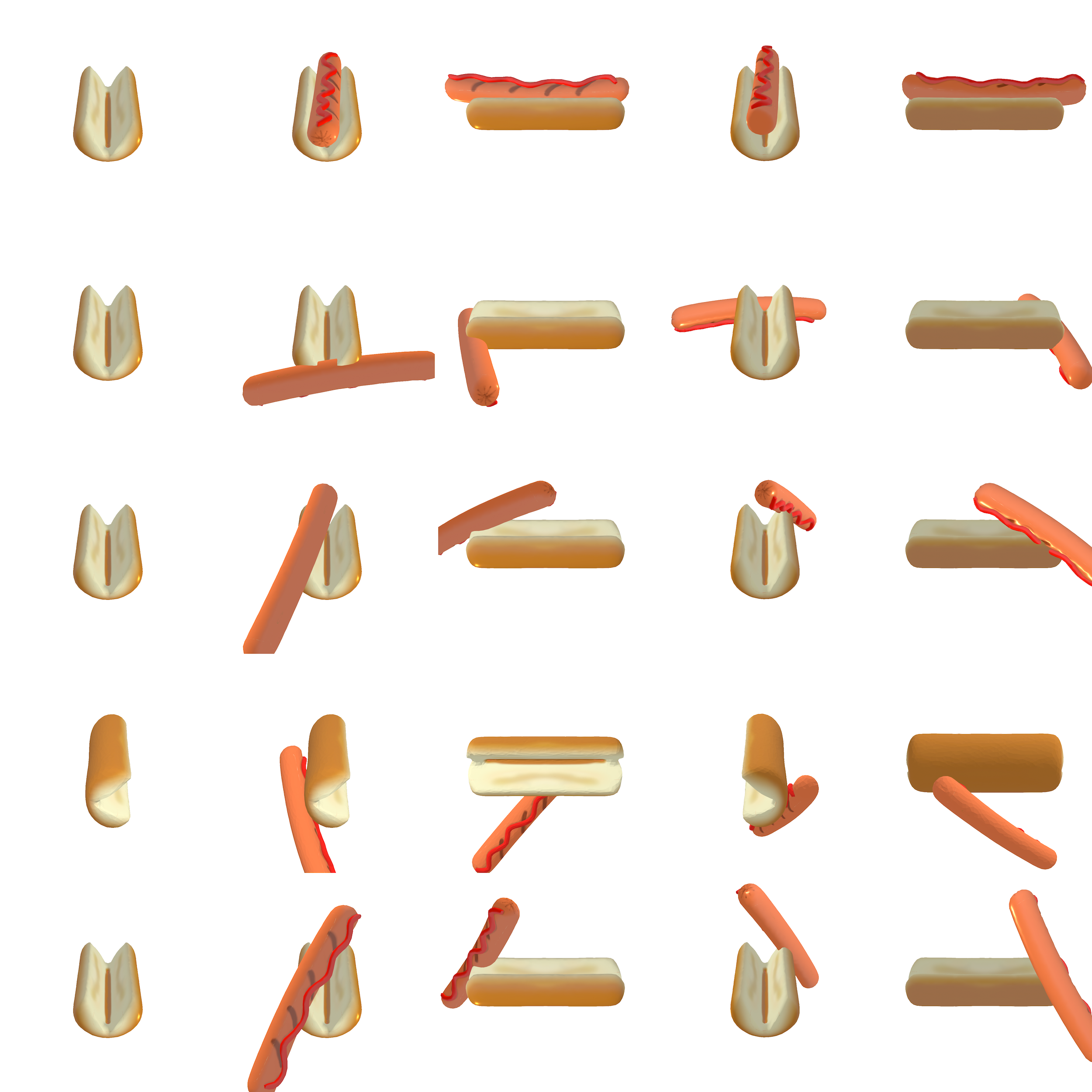}
  \end{minipage}
\caption{\textbf{Per-mesh qualitative comparison: hot dog and bun.}
Rows correspond to our method, B1, B2, SceneTeller, and SMC (top to bottom).
In each row, the leftmost panel shows the target mesh alone, followed by the final aligned meshes rendered from four different viewing angles.
Only our method produces a physically plausible configuration where the sausage rests inside the bun; B2 and SMC place the hot dog above the bun, but it floats without realistic contact.}
  \label{fig:hotdog_qual_supp}
\end{figure*}

\begin{figure*}[t]
  \centering
  {\large \textbf{``Man wearing a cowboy hat''}} \\[8pt]
  \begin{minipage}[t]{0.12\textwidth}
    \vspace{37pt} 
    \raggedleft
    \footnotesize
    \textbf{Ours} \\[74pt]
    \textbf{B1} \\[74pt]
    \textbf{B2} \\[74pt]
    \textbf{SceneTeller} \\[74pt]
    \textbf{SMC}
  \end{minipage}
  \hfill
  \begin{minipage}[t]{0.85\textwidth}
    \vspace{0pt}
    \includegraphics[width=\linewidth]{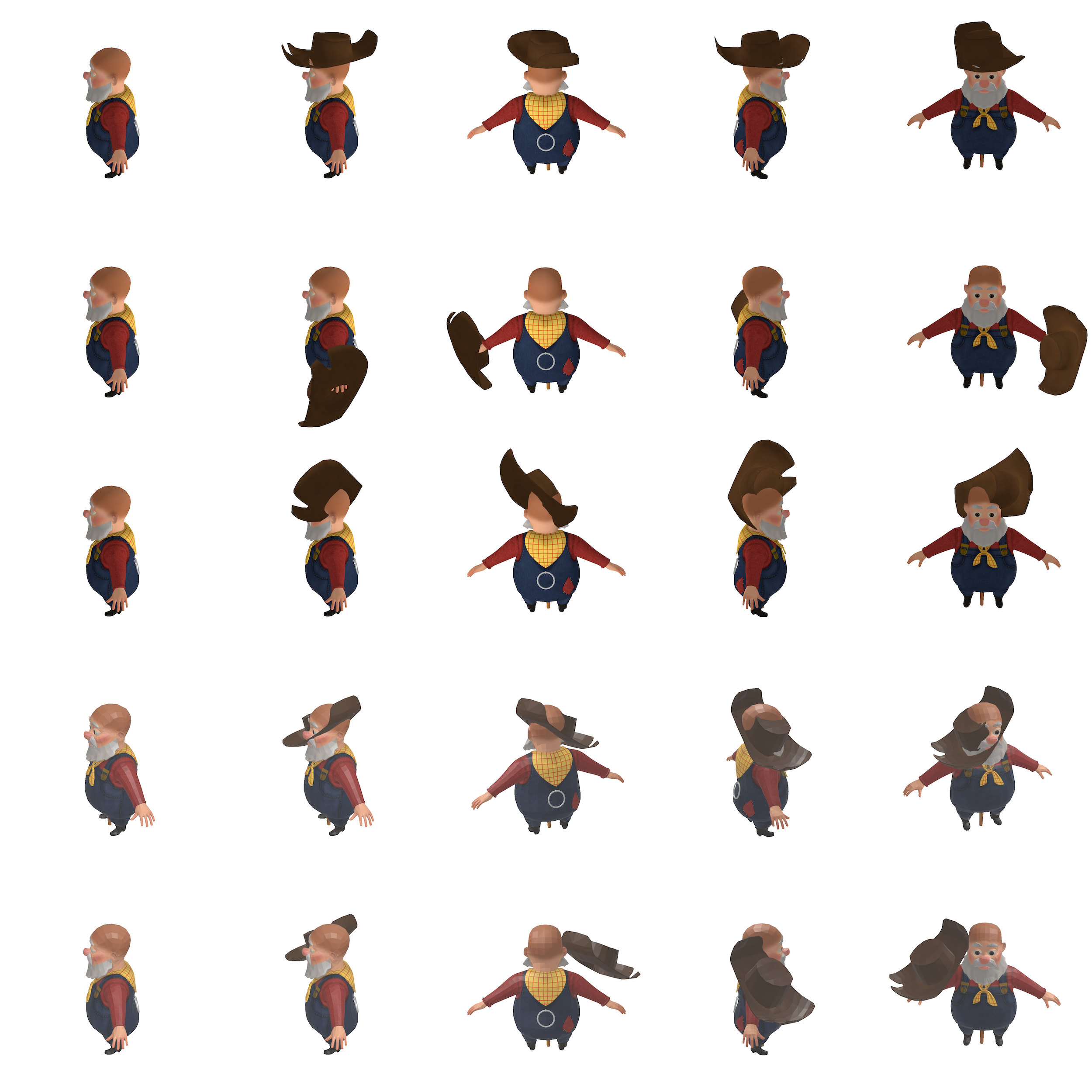}
  \end{minipage}
\caption{\textbf{Per-mesh qualitative comparison: man and cowboy hat.}
Rows correspond to our method, B1, B2, SceneTeller, and SMC (top to bottom).
In each row, the leftmost panel shows the target mesh alone, followed by the final aligned meshes rendered from four different viewing angles.
Our method positions the hat stably on the head; although B2, SMC, and SceneTeller place the hat close to the man's head, their configurations remain physically implausible, with poor contact or penetration.}
  \label{fig:cowboy_qual_supp}
\end{figure*}

\begin{figure*}[t]
  \centering
  {\large \textbf{``Chocolate box with a half-open lid on top''}} \\[8pt]
  \begin{minipage}[t]{0.12\textwidth}
    \vspace{37pt} 
    \raggedleft
    \footnotesize
    \textbf{Ours} \\[74pt]
    \textbf{B1} \\[74pt]
    \textbf{B2} \\[74pt]
    \textbf{SceneTeller} \\[74pt]
    \textbf{SMC}
  \end{minipage}
  \hfill
  \begin{minipage}[t]{0.85\textwidth}
    \vspace{0pt}
    \includegraphics[width=\linewidth]{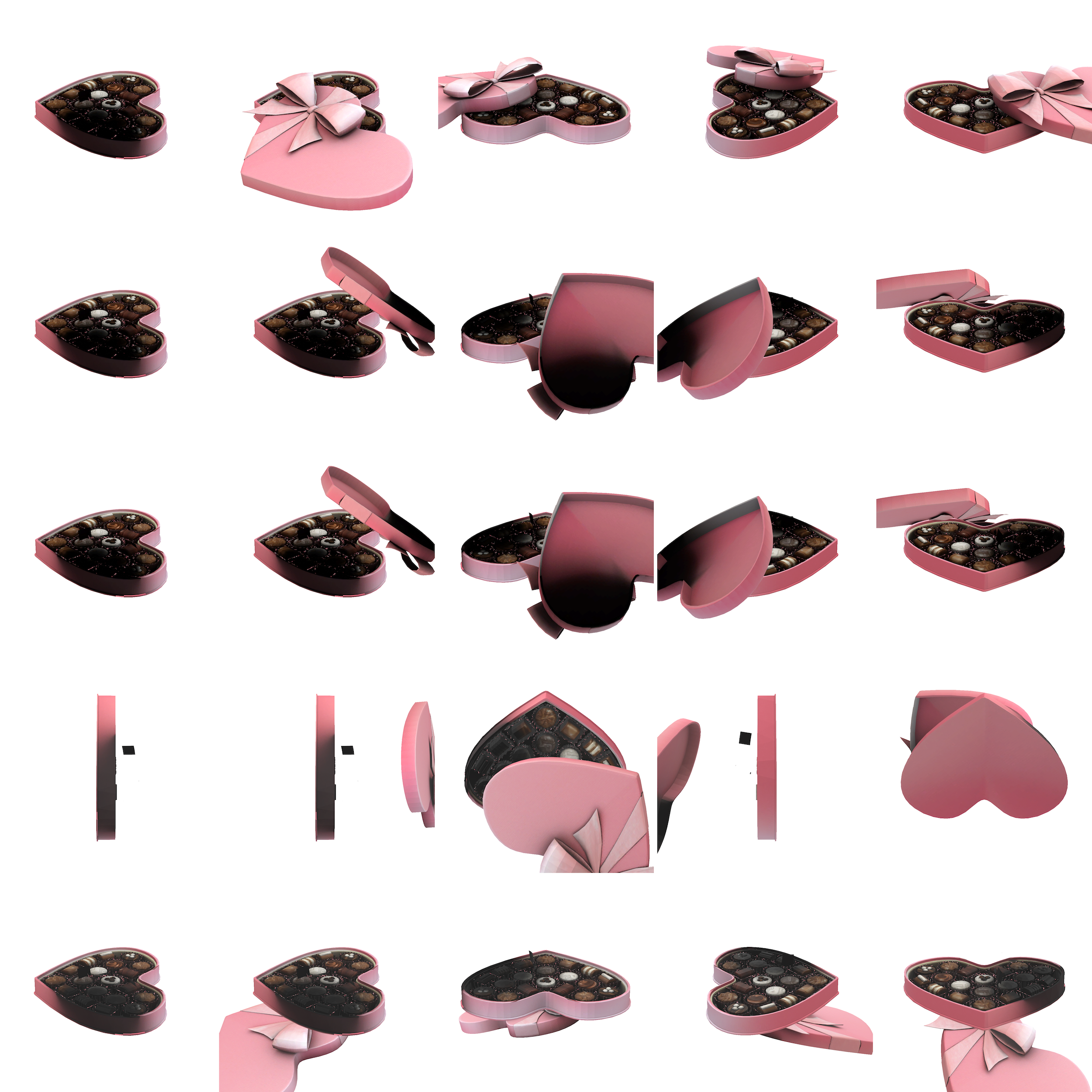}
  \end{minipage}
\caption{\textbf{Per-mesh qualitative comparison: chocolate box and lid.}
Rows correspond to our method, B1, B2, SceneTeller, and SMC (top to bottom).
In each row, the leftmost panel shows the target mesh alone, followed by the final aligned meshes rendered from four different viewing angles.
Only our result matches the prompt with a half-open lid resting on the box; in B1 and B2 the lid appears upside down, in SceneTeller it floats above the box, and in SMC it is placed below the box rather than on top.}
  \label{fig:chocolate_qual_supp}
\end{figure*}

\begin{figure*}[t]
  \centering
  {\large \textbf{``Judge’s gavel on its base''}} \\[8pt]
  \begin{minipage}[t]{0.12\textwidth}
    \vspace{37pt} 
    \raggedleft
    \footnotesize
    \textbf{Ours} \\[74pt]
    \textbf{B1} \\[74pt]
    \textbf{B2} \\[74pt]
    \textbf{SceneTeller} \\[74pt]
    \textbf{SMC}
  \end{minipage}
  \hfill
  \begin{minipage}[t]{0.85\textwidth}
    \vspace{0pt}
    \includegraphics[width=\linewidth]{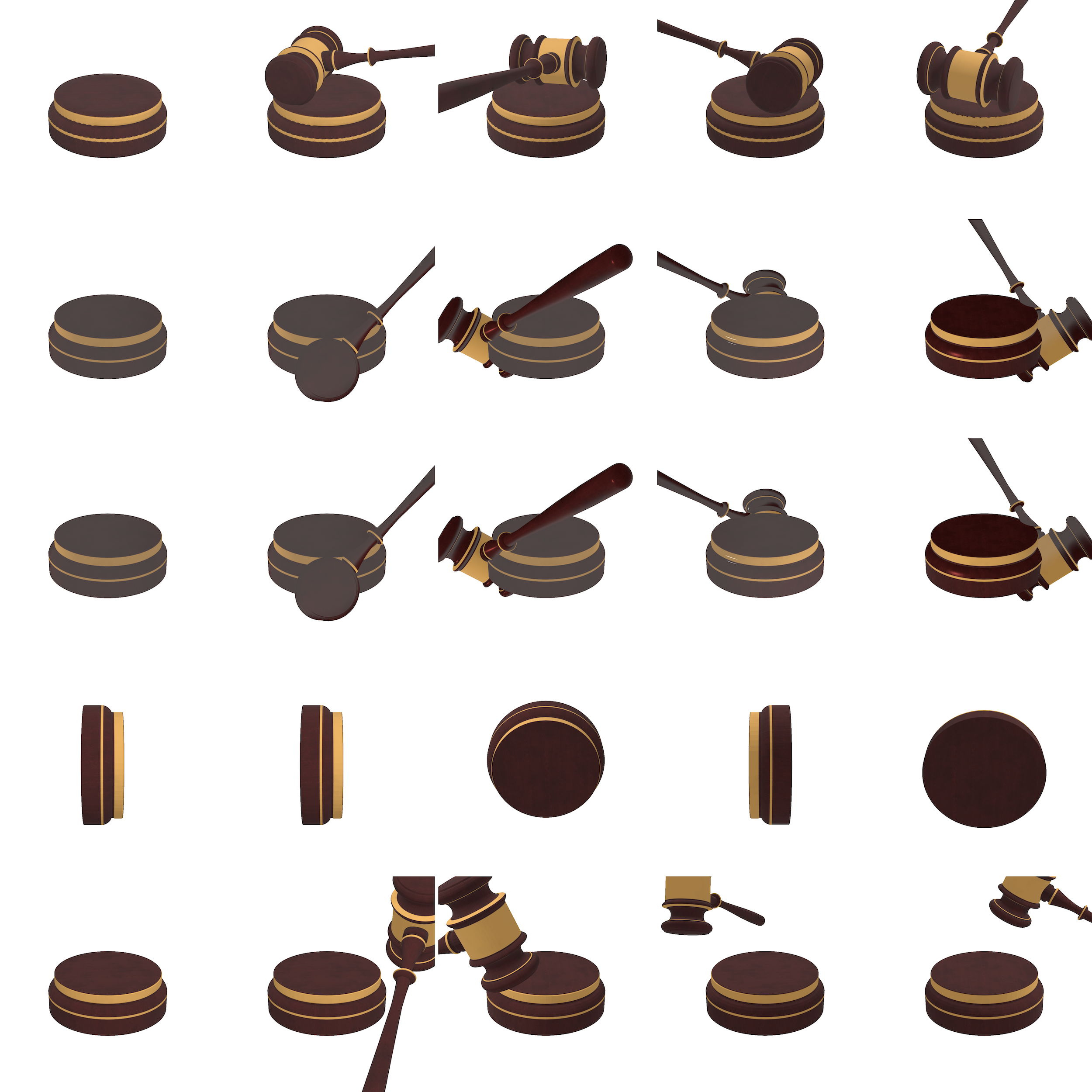}
  \end{minipage}
\caption{\textbf{Per-mesh qualitative comparison: judge’s gavel and base.}
Rows correspond to our method, B1, B2, SceneTeller, and SMC (top to bottom).
In each row, the leftmost panel shows the target mesh alone, followed by the final aligned meshes rendered from four different viewing angles.
Our method places the gavel directly and stably on its base; in B1 and B2 the gavel misses the base. Scene teller place the gavel far away from it's base, and in SMC the configuration can look text-aligned from some views but the gavel is not actually resting above the base.}
  \label{fig:judges_qual_supp}
\end{figure*}

\begin{figure*}[t]
  \centering
  {\large \textbf{``Toothpaste sits on top of toothbrush bristles''}} \\[8pt]
  \begin{minipage}[t]{0.12\textwidth}
    \vspace{37pt} 
    \raggedleft
    \footnotesize
    \textbf{Ours} \\[74pt]
    \textbf{B1} \\[74pt]
    \textbf{B2} \\[74pt]
    \textbf{SceneTeller} \\[74pt]
    \textbf{SMC}
  \end{minipage}
  \hfill
  \begin{minipage}[t]{0.85\textwidth}
    \vspace{0pt}
    \includegraphics[width=\linewidth]{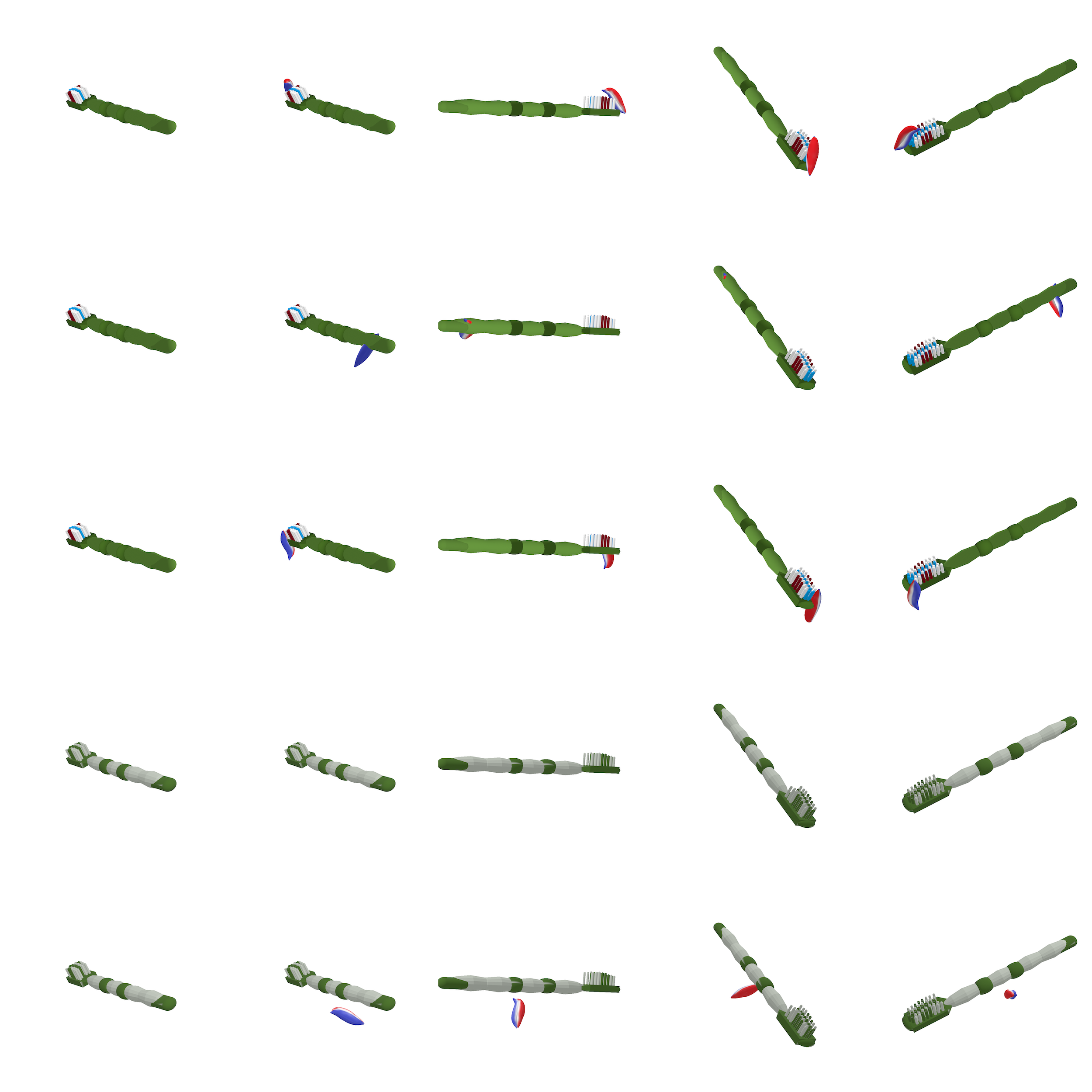}
  \end{minipage}
\caption{\textbf{Per-mesh qualitative comparison: toothpaste and toothbrush.}
Rows correspond to our method, B1, B2, SceneTeller, and SMC (top to bottom).
In each row, the leftmost panel shows the target mesh alone, followed by the final aligned meshes rendered from four different viewing angles.
While not perfect, our method yields a physically plausible placement of the toothpaste on the bristles. B2 comes close to placing it on top but still produces a floating, non-physical configuration.}
  \label{fig:toothpaste_qual_supp}
\end{figure*}

\begin{figure*}[t]
  \centering
  {\large \textbf{``Queen of hearts wearing a golden crown''}} \\[8pt]
  \begin{minipage}[t]{0.12\textwidth}
    \vspace{37pt} 
    \raggedleft
    \footnotesize
    \textbf{Ours} \\[74pt]
    \textbf{B1} \\[74pt]
    \textbf{B2} \\[74pt]
    \textbf{SceneTeller} \\[74pt]
    \textbf{SMC}
  \end{minipage}
  \hfill
  \begin{minipage}[t]{0.85\textwidth}
    \vspace{0pt}
    \includegraphics[width=\linewidth]{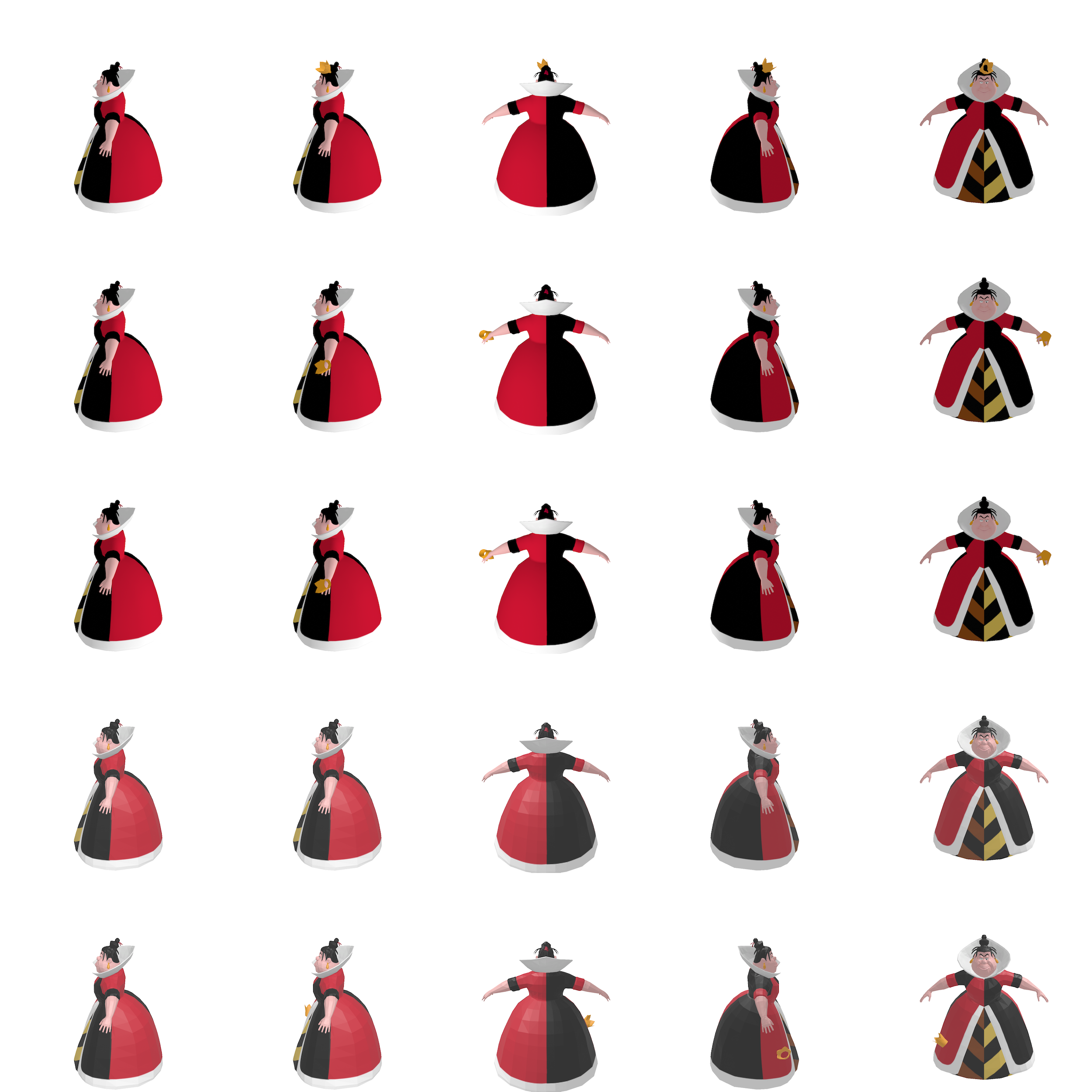}
  \end{minipage}
\caption{\textbf{Per-mesh qualitative comparison: queen and crown.}
Rows correspond to our method, B1, B2, SceneTeller, and SMC (top to bottom).
In each row, the leftmost panel shows the target mesh alone, followed by the final aligned meshes rendered from four different viewing angles.
In our result, the crown sits directly and stably on the queen’s head; in the other baselines the crown is either far from the head, or even not clearly visible in the final view.}
  \label{fig:queen_qual_supp}
\end{figure*}

We include additional per-mesh qualitative comparisons between our method and the four baselines (B1, B2, SceneTeller, and SMC).
For each prompt, the figures show (left to right) the target mesh alone followed by the final aligned meshes rendered from four different viewpoints.
These supplementary examples cover a diverse set of physical and semantic relationships and illustrate consistent improvements of our method across cases.
See 
\cref{fig:ice_cream_qual_supp,fig:blender_qual_supp,fig:hotdog_qual_supp,fig:cowboy_qual_supp,fig:chocolate_qual_supp,fig:judges_qual_supp,fig:toothpaste_qual_supp,fig:queen_qual_supp}
for the full set of comparisons.

\vspace{-1em}

\end{document}